\documentclass{emulateapj}

\usepackage{natbib}

\usepackage{graphicx}
\usepackage[space]{grffile}
\usepackage{latexsym}
\usepackage{amsfonts,amsmath,amssymb}
\usepackage{url}
\usepackage[utf8]{inputenc}
\usepackage{fancyref}
\usepackage{hyperref}
\hypersetup{colorlinks=false,pdfborder={0 0 0},}
\usepackage{textcomp}
\usepackage{longtable}
\usepackage{multirow,booktabs}

\newcommand{\pasa}{PASA}
\def\icarus{{Icarus}}

\usepackage{xspace}
\usepackage{amsmath}

\newcommand{\Rearth}{\ensuremath{R_{\oplus}}}

\newcommand{\water}{\ensuremath{ \textrm{H}_2\textrm{O}}\xspace}
\newcommand{\methane}{\ensuremath{ \textrm{CH}_4}\xspace}
\newcommand{\ammonia}{\ensuremath{ \textrm{NH}_3}\xspace}
\newcommand{\coo}{\ensuremath{ \textrm{CO}_2}\xspace}
\newcommand{\hh}{\ensuremath{ \textrm{H}_2}\xspace}

\shorttitle{Exoplanet Atmospheres}
\shortauthors{Crossfield}

\begin{document}


\title{Observations of Exoplanet Atmospheres}


\author{
Ian J. M. Crossfield\altaffilmark{1,2}}

\altaffiltext{1}{Lunar \& Planetary Laboratory, University of Arizona Lunar,
1629 E. University Blvd., Tucson, AZ, USA, \href{mailto:ianc@lpl.arizona.edu}{ianc@lpl.arizona.edu}}
\altaffiltext{2}{NASA Sagan Fellow}





\begin{abstract}
  Detailed characterization of an extrasolar planet's atmosphere
  provides the best hope for distinguishing the makeup of its outer
  layers, and the only hope for understanding the interplay between
  initial composition, chemistry, dynamics \& circulation, and
  disequilibrium processes.  In recent years, some areas have seen
  rapid progress while developments in others have come more slowly
  and/or have been hotly contested. This article gives an observer's
  perspective on the current understanding of extrasolar planet
  atmospheres prior to the considerable advances expected from the
  next generation of observing facilities.  Atmospheric processes of
  both transiting and directly-imaged planets are discussed, including
  molecular and atomic abundances, cloud properties, thermal
  structure, and planetary energy budgets.  In the future we can
  expect a continuing and accelerating stream of new discoveries,
  which will fuel the ongoing exoplanet revolution for many years to come.
\end{abstract}

\clearpage
\section{Introduction}

The exoplanet revolution is well underway.  The last decade has seen
order-of-magnitude increases in the number of known planets beyond the
Solar system. Surveys reveal that while hot Jupiters on few-day orbits
occur around $\lesssim$1\% of Sunlike stars
\citep{marcy:2005,howard:2012}; smaller, cooler sub-Neptunes and
super-Earths occur far more frequently \citep{dong:2013,fressin:2013},
and are 2--3$\times$ more common still around low-mass stars
\citep{bonfils:2013,mulders:2015a,dressing:2015}.  At larger orbital
separations planets with super-Jovian masses are rare, occurring
around $<$5\% of stars \citep{biller:2013a}.

Even greater than the advances made in planetary demographics is the
dramatic progress made toward understanding the atmospheres of these
distant worlds.  Detailed characterization of a planet's atmosphere
provides the best hope for distinguishing the makeup of its outer
layers, and the {\em only} hope for understanding the interplay
between initial composition \citep{madhusudhan:2011,mordasini:2012b},
chemistry \citep{fortney:2010,barman:2011}, dynamics \& circulation
\citep{showman:2009}, and disequilibrium processes
\citep{moses:2013,line:2013b}.  Mass and radius measurements alone
cannot uniquely identify the H$_2$, ice, and rock content of many
planets \citep{adams:2008}.  Atmospheric measurements are essential.

Given the rapid expansion of the field, a steady stream of updated
reviews is required to keep up to date.  A number of excellent review
articles have been published on the topic of exoplanet atmospheres,
and much in these remains accurate and relevant
\citep{marley:2007ppv,burrows:2010,seager:2010review,madhusudhan:2014,burrows:2014,bailey:2014}. Since
the latest of these do not discuss the most recent exciting
developments, while others discuss results now seen to have been illusory,
the time is ripe for another  comprehensive review.

This article gives one observer's perspective  of
our current understanding of extrasolar planet atmospheres. The document is
organized as follows: observing methods and the modeling approaches
used to infer atmospheric properties are described in
Secs.~\ref{sec:obs} and~\ref{sec:retrieval}, respectively.
Sec.~\ref{sec:composition} summarizes the status of measurements of
atmospheric composition and chemistry, leading in to a discussion of
clouds and hazes in Sec.~\ref{sec:clouds} and circulation, energy
budgets, and variability in Sec.~\ref{sec:energy}.  Finally,
Sec.~\ref{sec:future} concludes with a look toward the exciting future
of the field.

\section{Exoplanet Observing  Techniques}
\label{sec:obs}
\begin{figure*}[tb]
\begin{center}
\includegraphics[width=7in]{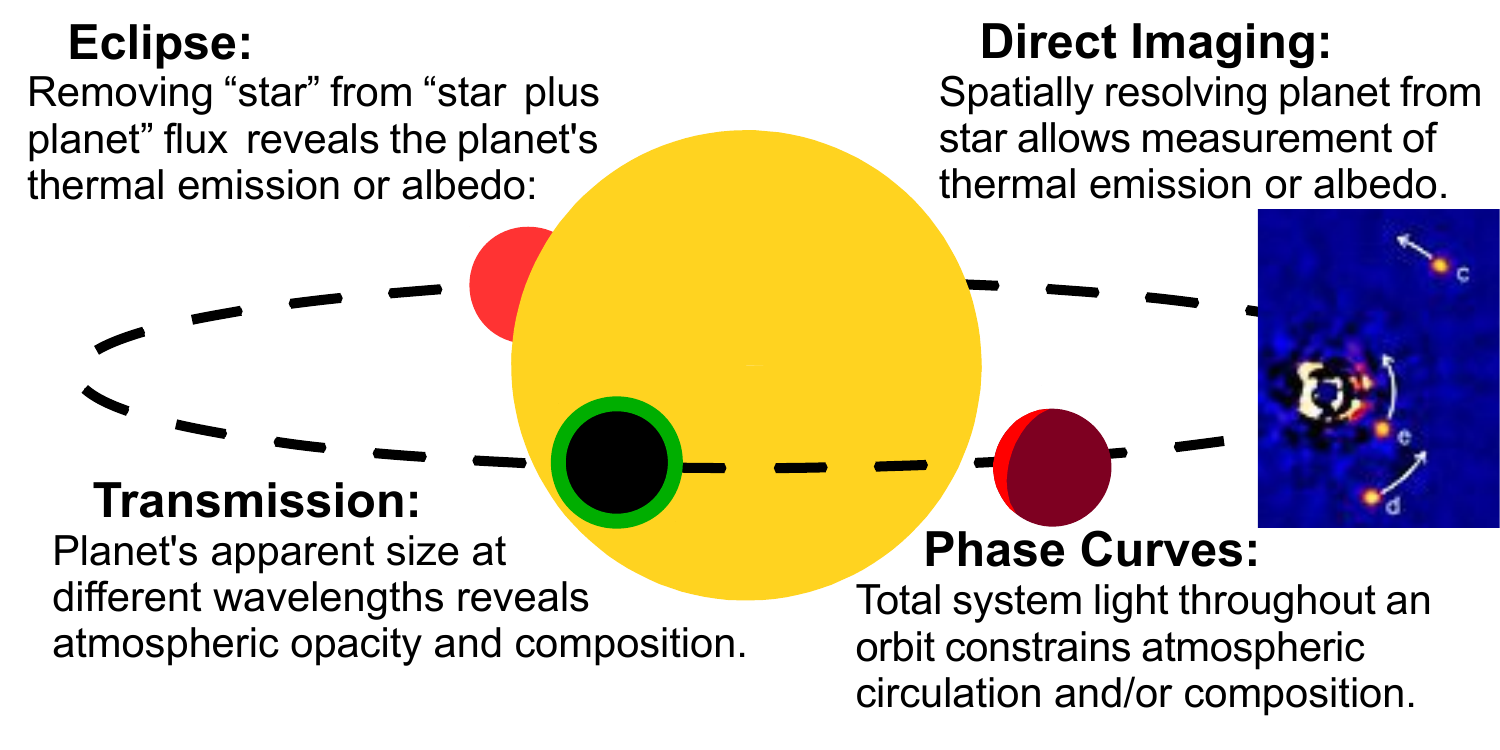}
\caption{\label{fig:methods} Observational approaches used to
  characterize the atmospheres of extrasolar planets. 
}
\end{center}
\end{figure*}

Before diving in to the many recent advances in the field, it is
worthwhile to review the various observational techniques used to
obtain the necessary data.  This section begins with direct imaging in
Sec.~\ref{sec:direct}, then moves to shorter-period (typically
transiting) planets: occultation studies in Sec.~\ref{sec:transit},
phase curves in Sec.~\ref{sec:phase}, and high-dispersion spectroscopy
in Sec.~\ref{sec:hds}.  Fig.~\ref{fig:methods} schematically depicts
each of these approaches.

\subsection{Direct imaging}
\label{sec:direct}
\begin{figure}[tb]
\includegraphics[width=3.5in]{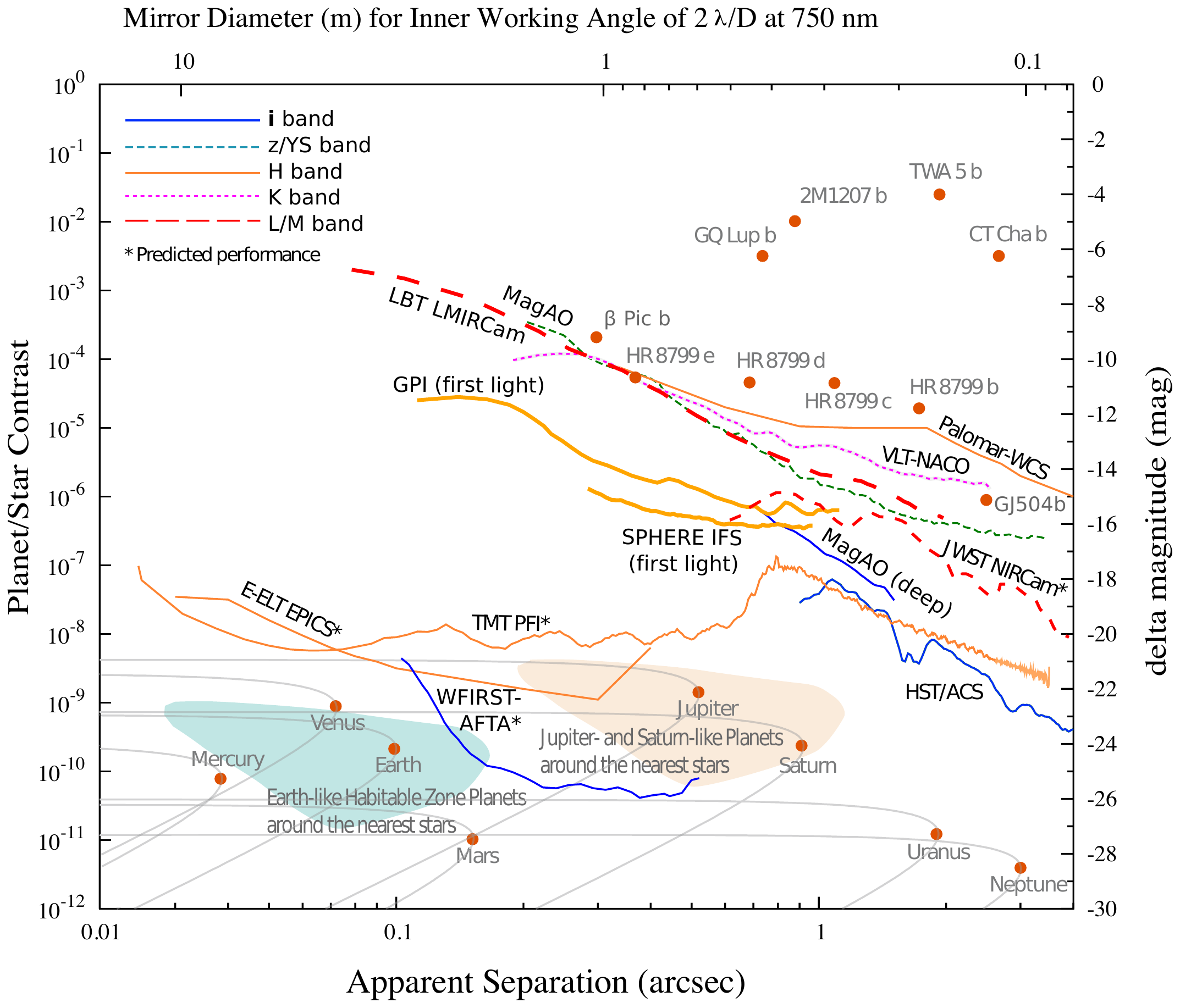}
\caption{\label{fig:contrast} Planet/star contrast
  achieved via current and planned direct imaging instruments. Orange
  points indicate a subset of detected planets (at upper right, as
  seen in $K$ band), and Solar System planets at 10~pc (bottom,
  assuming reflected light). Updated from \cite{mawet:2012}, with additional data added
  \citep[][Males et al., in prep., and private communication from A.\ Vigan]{close:2014,macintosh:2014,kuzuhara:2013,skemer:2014ao,spergel:2015}. }

\end{figure}

Direct imaging --- spatially resolving a planet at some angular
separation from its host star --- is a challenging endeavor. At a
distance of 10~pc a planet in a Jupiter-like orbit ($a=5$~AU) is just
0.5'' from its host star, well within the $\sim$1'' angular resolution
available to most ground-based telescopes. Only with adaptive optics
\citep{hardy:1998} can these facilities reach their theoretical,
diffraction-limited angular resolution of 0.2''$(\lambda/\micron)
(D/m)^{-1}$.  AO-equipped or space-based telescopes can therefore
resolve extrasolar planets on wide orbits, but even for young, hot
planets still glowing from the heat of their formation the planets are
at best $\lesssim10^{-4}$ times as bright as the nearby host
star. Adaptive optics or space-based facilities are essential to
produce diffraction-limited images, but even fairly faint levels of
residual scattered starlight must be removed before any faint planet
can be seen. Fig~\ref{fig:contrast} shows the current state of the art
relative to a variety of known planets.

The art and science of direct imaging focuses to a considerable degree
on removing the residual starlight as effectively as possible.  Steady
progress has been made since the first detections
\citep{chauvin:2004,marois:2008}. A variety of techniques including
simultaneous differential imaging
\citep[][]{smith:1987,rosenthal:1996}, spectral differential
imaging \citep[][]{sparks:2002}, angular differential imaging
\citep[][]{marois:2006}, and others have been developed; see
\cite{mawet:2012} for a review. Whatever the approach, observations
taken with either successive photometric filters or a spectrograph
then measure the planet's thermal emission.  In principle reflected
light could also be detected from these planets, but the planet-star
contrast ratios are much lower: $(2\times 10^{-7}) A_G (R_p/R_J)^2
(a/\textrm{AU})^{-2}$, where $A_G$ is the planet's geometric
albedo. Nonetheless future ground- and space-based telescopes will
study some small subset of giant planets in reflected light
\citep{crossfield:2013a,quanz:2015,marley:2014,burrows:2014b,hu:2014b}.

Similar to directly-imaged planets but easier to study are solivagant
(free-floating) or widely-separated planetary-mass objects. Whether
these objects form like traditional planets is as yet undetermined
\citep[see the discussion in][]{beichman:2014b}, but it is clear that
these wanderers can also provide key insights into atmospheric
properties of cool planetary atmospheres. These have no nearby host
star and so can be characterized even by seeing-limited observations
(without the need for adaptive optics), just as brown dwarfs have been
studied for several decades. A further advantage of any type of direct
imaging is that once the necessary contrast has been reached, one
obtains much higher S/N and spectral resolution than is possible for a
transiting planet of comparable brightness. This is because photon
shot noise is much reduced by suppressing the host starlight and
planets are visible all night, every night rather than for just a few
hours per orbit.

\subsection{Transits and Eclipses}
\label{sec:transit}


Transiting planets occult their host stars, resulting in a periodic
flux decrement approximately equal to $(R_P/R_*)^2$ -- this is
$10^{-4}$ for Earth orbiting the Sun, but is $2 \times 10^{-3}$ for a
1.5\,\Rearth super-Earth orbiting an M4 dwarf and 1--2\%\ for a
typical hot Jupiter.  When the planet has an atmosphere, its
atmospheric opacity varies with wavelength and so the radius at which
the planet's atmosphere becomes optically thick changes with
wavelength \citep{seager:2000,brown:2001,hubbard:2001}. One therefore
measures the flux decrement at multiple wavelengths; the observed
quantity $(R_P(\lambda)/R_*)^2$ is termed the ``transmission
spectrum.''  At moderate spectral resolution, its spectral features
have amplitudes of a few times $H R_P/R_*^2$
\citep{miller-ricci:2009}, where $H$ is the atmospheric scale
height. Thus to measure transmission features in an Earth-Sun twin
system requires a fractional precision approaching $10^{-6}$ (1~ppm), but this
increases to $\sim$15~ppm for our \hh-rich super-Earth orbiting its M
star and even larger for hot Jupiters.  \citep[In fact, both S/N
considerations and refractive effects limit the useful study of
habitable atmospheres in transit to M star
systems;][]{betremieux:2014}.  Clouds and hazes tend to mute spectral
features from molecular and atomic absorbers, producing nearly flat
transmission spectra \citep{seager:2000,marley:2013}.

Transiting planets on circular orbits also pass behind their host
stars, blocking the planets' thermal emission and decreasing the flux
observed from the systems by $(R_P/R_*)^2 F_{\nu,P}/F_{\nu,*}$, where
$F_\nu$ is the emergent flux density of the planet or star. At longer
wavelengths $F_{\nu,P}$ is dominated by thermal emission resulting from the
planet's internal heat and/or reprocessed starlight; at
shorter wavelengths scattered starlight may also be a significant
contribution.  Assuming simple, blackbody-like emission spectra, at
15\,\micron\ the eclipse signal is just 0.8~ppm for an Earth-Sun twin
system, $\sim$30~ppm for the super-Earth orbiting an M dwarf described
above, and a few times $10^{-3}$ for hot Jupiters.  To characterize
atmospheric features rather than merely detect planetary emission
requires perhaps a further tenfold improvement in precision. Planetary
emission or reflection can also be studied via phase curves
(Sec.~\ref{sec:phase}). Fig.~\ref{fig:methods} demonstrates how
transits, eclipses, and phase curves all provide complementary
insights into a transiting planet's atmosphere.

Most observations to date have employed low-dispersion spectroscopy or
broadband photometry. An alternative but less-explored method of
transmission observations, using high-dispersion spectroscopy from
large-aperture telescopes, is discussed below in
Sec.~\ref{sec:hds}. The most profitable tool for studying transiting
exoplanet atmospheres has been space-based spectroscopy with HST,
which has been used to study the atmospheres of a rapidly-growing
number of both hot Jupiters and sub-Jovians
\citep{charbonneau:2002,lecavelier:2008haze189,evans:2013,kreidberg:2014,knutson:2014a}.
Observations with Spitzer and ground-based telescopes have also been
frequently employed \citep{knutson:2008,desert:2011,croll:2011a} but
with typically less secure characterization of atmospheric properties.
{JWST will soon be the superlative instrument for these studies,}
providing over 10$\times$ the total spectral range of HST/WFC3, with a
sensitivity $\sim$3$\times$ greater than HST and 8$\times$ greater
than Spitzer \citep{beichman:2015}.


\subsection{Phase curves}
\label{sec:phase}
Phase curves track the modulation of planetary thermal emission and/or
reflected starlight throughout a planet's orbit. These observations
complement secondary eclipses, which only measure emission from a
transiting planet's day side.  Phase curves provide a wealth of
information about planetary atmospheric dynamics and energetics by
measuring longitudinal brightness temperature maps, thereby
constraining atmospheric conditions across the planet's surface. The
phase curve amplitude determines the day-night temperature contrast in
the thermal infrared, or constrains the planet's albedo at shorter
wavelengths.  The phase offset measures the longitude of the planet's
brightest or dimmest point relative to the substellar meridian.

Phase curve morphologies are in general a degenerate
outcome an atmosphere's radiative, advective, and drag timescales
and the efficiency with which winds circulate incident stellar
energy around the planet.
If the incident stellar flux is immediately re-radiated, the
hottest point on the planet is at the substellar point (which receives
the highest level of incident flux).  Nonzero infrared phase offsets thus imply
heat transport around the planet.
Clouds or hazes can move the infrared photosphere to higher altitude,
resulting in a larger observed temperature contrast and smaller phase
offset \citep[][]{pont:2013}.  Large phase offsets could indicate
strong advection or drag forces \citep{cowan:2011circ,perezbecker:2013}
or re-radiation of thermal energy deposited by shock fronts in the
planet's atmosphere where supersonic equatorial jets transition to
subsonic speeds \citep{heng:2012c}.  Shocks and supersonic flows may
be absent if magnetic drag forces are significant. Non-negligible
Lorentz forces may occur in the presence of planetary magnetic fields
when alkali atoms are ionized
\citep[][]{perna:2010a,rauscher:2013}. 

To obtain a phase curve, one observes a system for some substantial
fraction of an orbit \citep{knutson:2007b}. If the planet is
transiting, one or more secondary eclipses provides an absolute
reference to disentangle stellar flux from combined ``star plus
planet'' light. If the planet does not transit, the resulting
constraints on atmospheric properties are looser
\citep{cowan:2007,crossfield:2010}.  While a phase curve provides
substantially more information than a secondary eclipse, the former
requires considerably more observing time and is more susceptible to
degradation by slowly-varying, low-amplitude instrumental systematics
\citep{agol:2010,crossfield:2012b}.

\begin{figure*}[tb]
\centering
\includegraphics[width=6in]{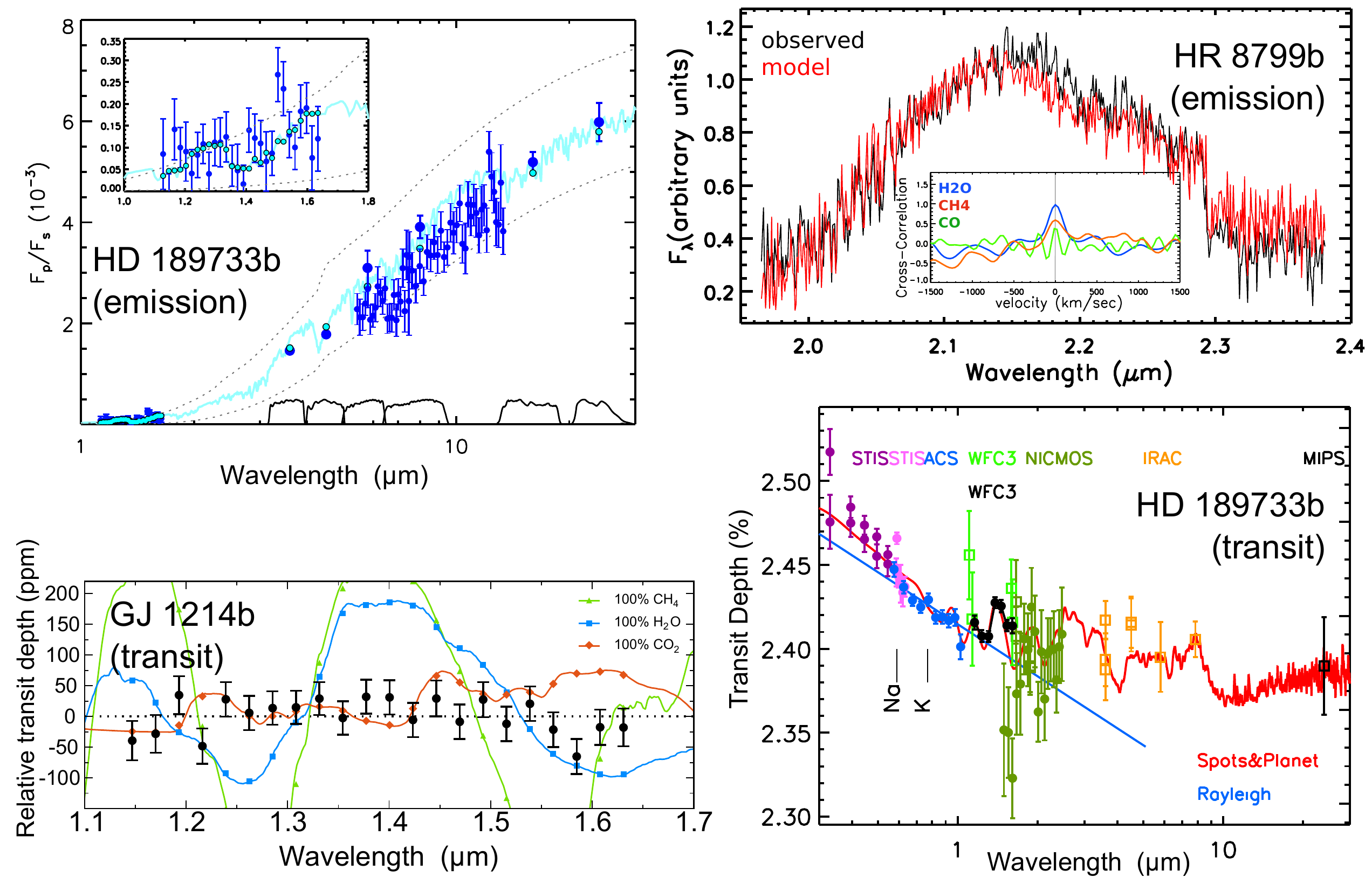}
\caption{\label{fig:spectra} Some of the best-characterized
  exoplanets, highlighting the main atmospheric conclusions to
  date. Clockwise from upper left: approximately isothermal emission spectra of
  hot Jupiters with weak \water\ absorption features (HD~189733b,
  emission); high-quality emission spectra of directly imaged planets,
  revealing \water, \methane, and CO (HR~8799b); transmission spectra
  of hot Jupiters showing short-wavelength slopes, \water\ at
  1.4\,\micron, and no conclusive features at longer wavelengths
  (HD~189733b, transit); and flat, featureless transmission spectra of
  cool sub-Jovians, indicative of high-altitude haze
  (GJ~1214b). Figures adapted from
  \cite{crouzet:2014,todorov:2014,barman:2015,mccullough:2014,kreidberg:2014}.
}

\end{figure*}

\subsection{High-dispersion spectroscopy}
\label{sec:hds}

Ground-based telescopes permit large, massive spectrographs with
$\lambda/\Delta \lambda \gtrsim 30,000$, vs.\ a few thousand at most
for HST or JWST.  The increased spectral resolution dramatically increases the
contrast of absorption features in exoplanetary spectra, more than making up
for the increased calibration challenges resulting from ground-based
spectroscopy.  

When applied to short-period hot Jupiters, high-dispersion
spectroscopy treats planet-star systems as double-lined spectroscopic
binaries. The technique now regularly detects thermal emission from
hot Jupiters at the $\sim 10^{-4}$ level relative to the star,
ascertains whether targets have a temperature inversion, and measures
molecular abundances as small as $\sim10^{-5}-10^{-4}$.  This powerful
new technique has produced a surge in the number of planets with
robust molecular detections or constraints on atmospheric properties
\citep{snellen:2010,crossfield:2011,brogi:2012,brogi:2013,brogi:2014,rodler:2012,rodler:2013b,birkby:2013,dekok:2013,lockwood:2014}.
A further strength of high-dispersion Doppler spectroscopy is its
applicability to non-transiting planetary systems, where
high-dispersion observations break the $\sin i$ degeneracy that
plagues interpretation of non-transiting phase variations.

When coupled to an AO system, high-dispersion spectroscopy can also be
use to great effect to study directly imaged planets. This has been
done for several planets at medium resolution
\citep[][]{konopacky:2013,barman:2015} and at $\lambda/\Delta \lambda
\sim 90,000$; for beta~Pic~b \citep{snellen:2014}. The latter type of
observation directly constrains planetary rotation rates and could
eventually place high-fidelity constraints on planets' thermal
structure \citep{line:2014c,line:2015} and global 2D surface brightness profiles
\citep{crossfield:2014b}.

\section{Inferring Atmospheric Properties from Observations}
\label{sec:retrieval}

The previous section describes the observations used to constrain an
exoplanet's atmosphere, but these data do not automatically provide
atmospheric parameters. Only by comparison to physically relevant
models can the parameters of interest --- temperature, abundances,
etc. --- be inferred. Like the observations themselves, obtaining
useful model-derived properties is challenging. In the simplest cases,
toy models provide useful insights --- constraining temperatures via
comparison with blackbodies \citep{charbonneau:2005,deming:2005a}, or
deriving temperature and mean molecular weight from Rayleigh scattering
slopes \citep{lecavelier:2008haze209} --- but typically more elaborate
models are needed.

Analyses of exoplanet atmospheres fall into two general categories:
forward models, which assume a particular combination of parameters to
generate an observable (e.g., a spectrum); and the inverse or
``retrieval'' approach, which explicitly aims to determine the
best-fit parameters and their uncertainties from the available
data. The recent review by \cite{madhusudhan:2014} discusses both of
these techniques.  Though sometimes seen as at odds with each other,
the two approaches are complementary.

Forward models are essential to predict the observable properties of
previously unobserved objects for which scant data exists. Such models
have played a key role in focusing attention on especially interesting
or easily observable atmospheric features: e.g., alkali absorption in
hot Jupiters \citep{seager:2000,barman:2002,charbonneau:2002}. Forward models also
often include more detailed treatments of atmospheric properties which
would be too time-consuming to include in a retrieval; for example,
dust formation and dispersal in substellar atmospheres
\citep{helling:2008}.

Retrieval tools \citep[][Cubillos et al., in
prep.]{irwin:2008,madhusudhan:2009,benneke:2012,line:2013} necessarily
include some form of forward model, incorporated within an algorithm
that explores the atmospheric parameter phase space consistent with
the measurements in hand. To adequately sample the atmospheric
parameter posterior distributions in a timely manner various
simplifications are often made: molecular abundances may be constant
with altitude, thermal profiles may assume an arbitrary
parametrization without enforcement of radiative equilibrium, and so
forth. However, techniques and computing power are quickly advancing
to the point where many of these assumptions can already be relaxed
\citep{line:2014c,line:2015,benneke:2015}. In any case, the power of retrieval
is that --- subject to the model assumptions and the reliability of
the data --- the technique provides statistically robust confidence
intervals on the desired parameters.

For whatever reason, retrieval analyses have focused mainly on
transiting planets \citep[with some exceptions;][]{lee:2013} while
analyses of directly imaged planets have relied mainly on forward
models. Some of the latter studies compare observations to
multidimensional grids of precomputed template spectra, though the
variables are typically fewer than in the retrieval analyses described
above. It seems likely that the field could benefit from the
complementary perspectives that would be provided by application of
retrieval methods to spectra of directly imaged planets.

%


\section{Molecules \& Atoms}
\label{sec:composition}
One of the first questions asked about an exoplanet's atmosphere is,
``What is it made of?''  The last few years have seen rapid strides
toward answering this question for a wide range of molecules and
atomic species, as were seen in the preceding years for studies of
brown dwarfs \citep[e.g.,][]{lodders:2006}.  This section first
considers molecular detections from both spectroscopy and broadband
photometry, and then discusses recent measurements of atoms and ions
in planets' upper atmospheres. A comprehensive overview of recent
detections is also given by \cite{bailey:2014}, and \cite{heng:2015c}
present an excellent discussion of the key concepts involved in
atmospheric chemistry. Clouds and hazes are subsequently discussed in
Sec.~\ref{sec:clouds}, and thermal structure in Sec.~\ref{sec:inv}.

\subsection{Molecular Detections and Abundances}
\begin{figure*}[tb]
\begin{center}
\includegraphics[width=3.3in]{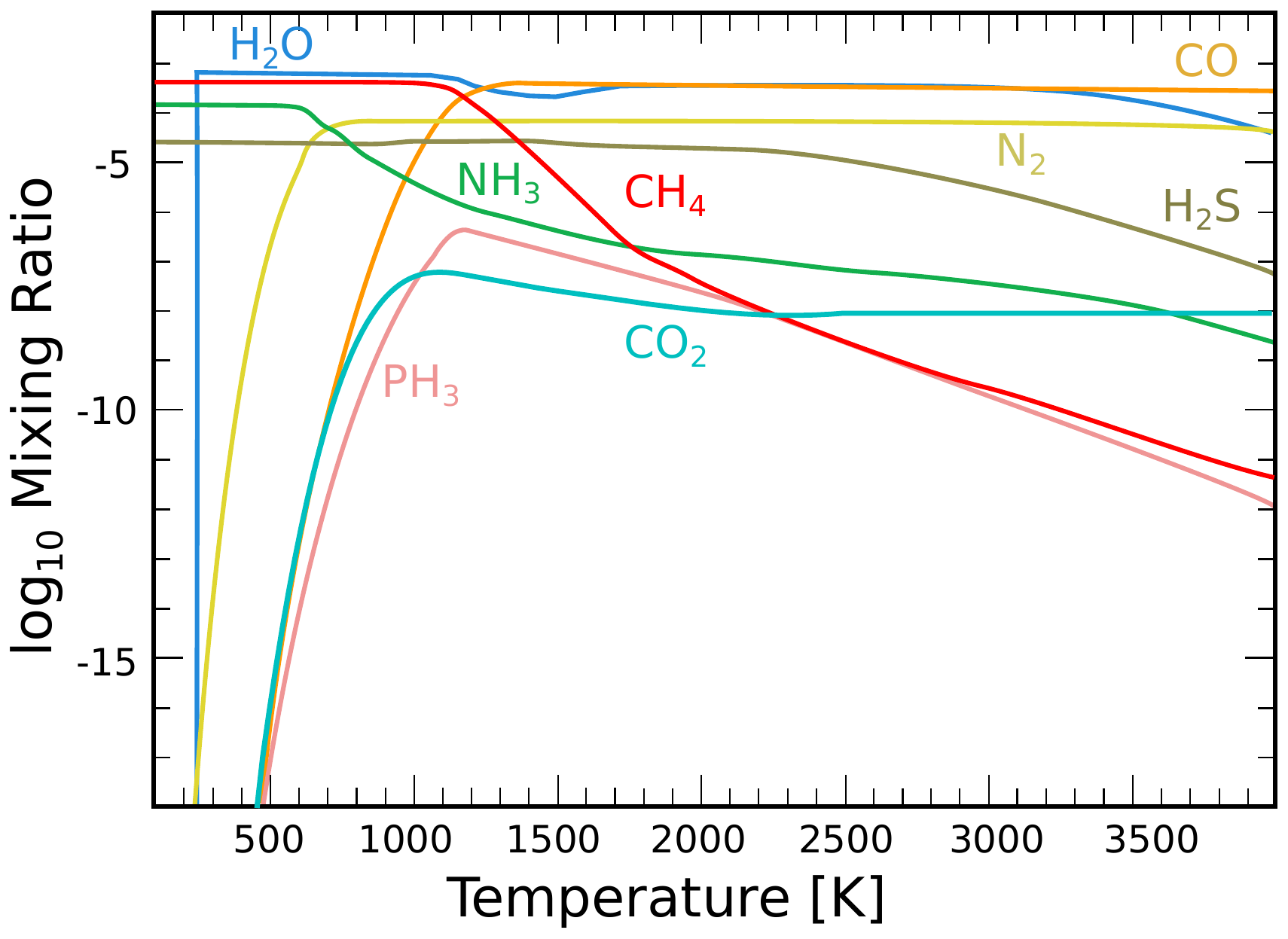}
\includegraphics[width=3.0in]{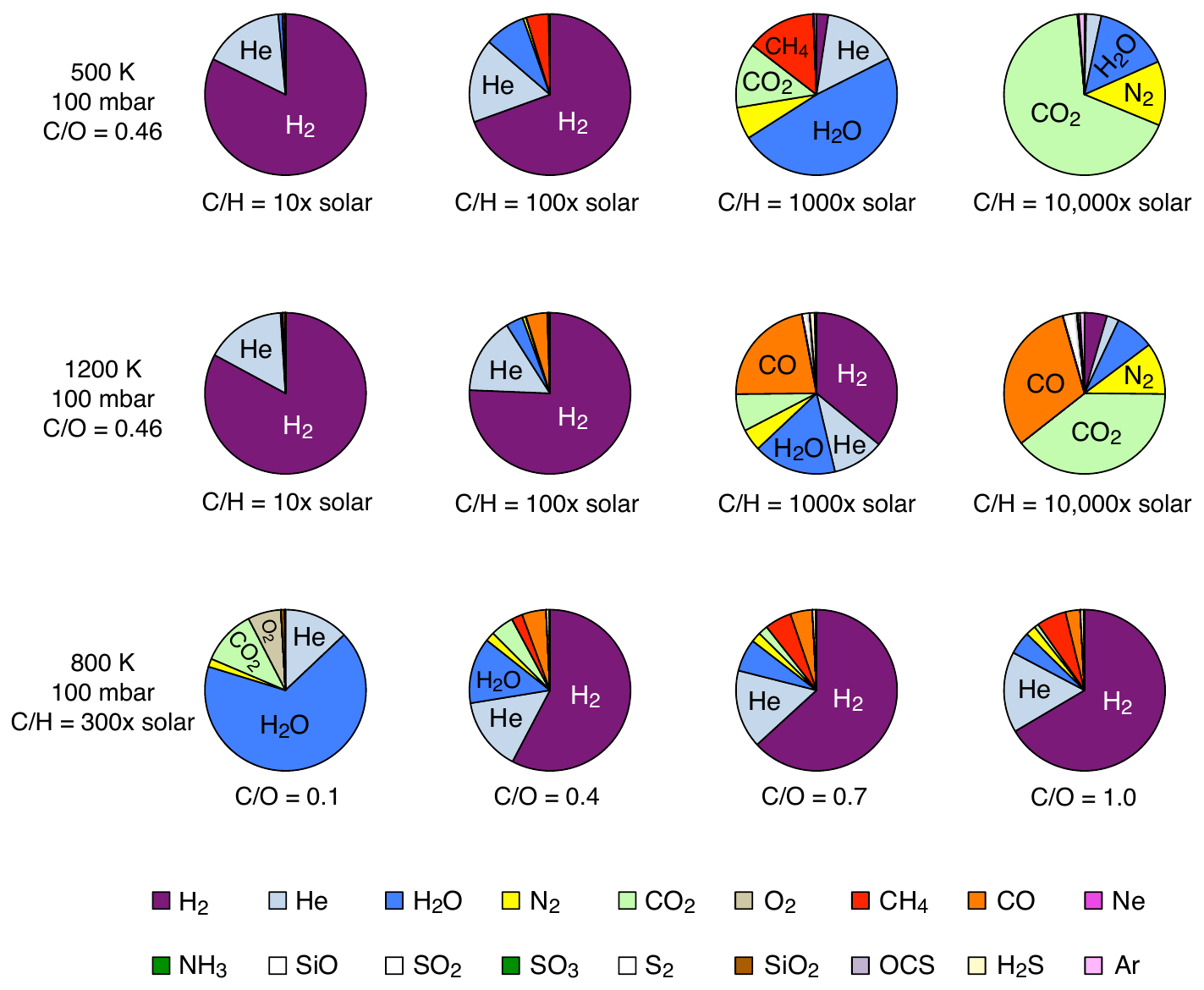}
\caption{\label{fig:eq} Molecular equilibrium abundances: {\em Left}:
  vs.\ temperature in a Solar-metallicity atmosphere at a total
  pressure of 1~bar; rainout, photochemistry, quenching, etc.\ are all
  neglected \citep[adapted from][\citeauthor{miguel:2014}
  2014]{sharp:2007}. H$_2$ would be above the top of the plot.
  {\em Right:} For generic hot Neptunes with a range of temperatures,
  metallicities, and C/O ratios \citep[from][]{moses:2013}. }
\end{center}
\end{figure*}

Planetary atmospheres are composed primarily of molecules. The
relative abundances of different molecular species depend on the
complex interplay between bulk chemical composition resulting from the
planet's formation \& evolution, differentiation, atmospheric
circulation, and photochemical and other disequilibrium processes.
Fig.~\ref{fig:eq} shows the expected abundances of various molecules
in equilibrium (but see Sec.~\ref{sec:diseq}).  For typical,
solar-like abundances the 1000--2500~K temperatures of most giant
exoplanets studied to date indicates that CO and \water\ will be the
most abundant and spectroscopically active species in equilibrium
conditions \citep{burrows:1999,lodders:2006}.  \coo\ will also be present in high
metallicity environments at these warmer temperatures; its abundance
remains lower than that of CO for all but the highest metallicities
\citep[e.g.,][]{moses:2013}. Below $\sim$1300~K \methane\ should
become more abundant than CO, and below $\sim$700~K \ammonia\ become
increasingly abundant as well \citep{burrows:1999}.  The following
sections review the measurements of various molecules detected by
either spectroscopy (Sec.~\ref{sec:molspec}) or broadband photometry
(Sec.~\ref{sec:molphot}).

\begin{figure}[tb]
\begin{center}
\includegraphics[width=3.5in]{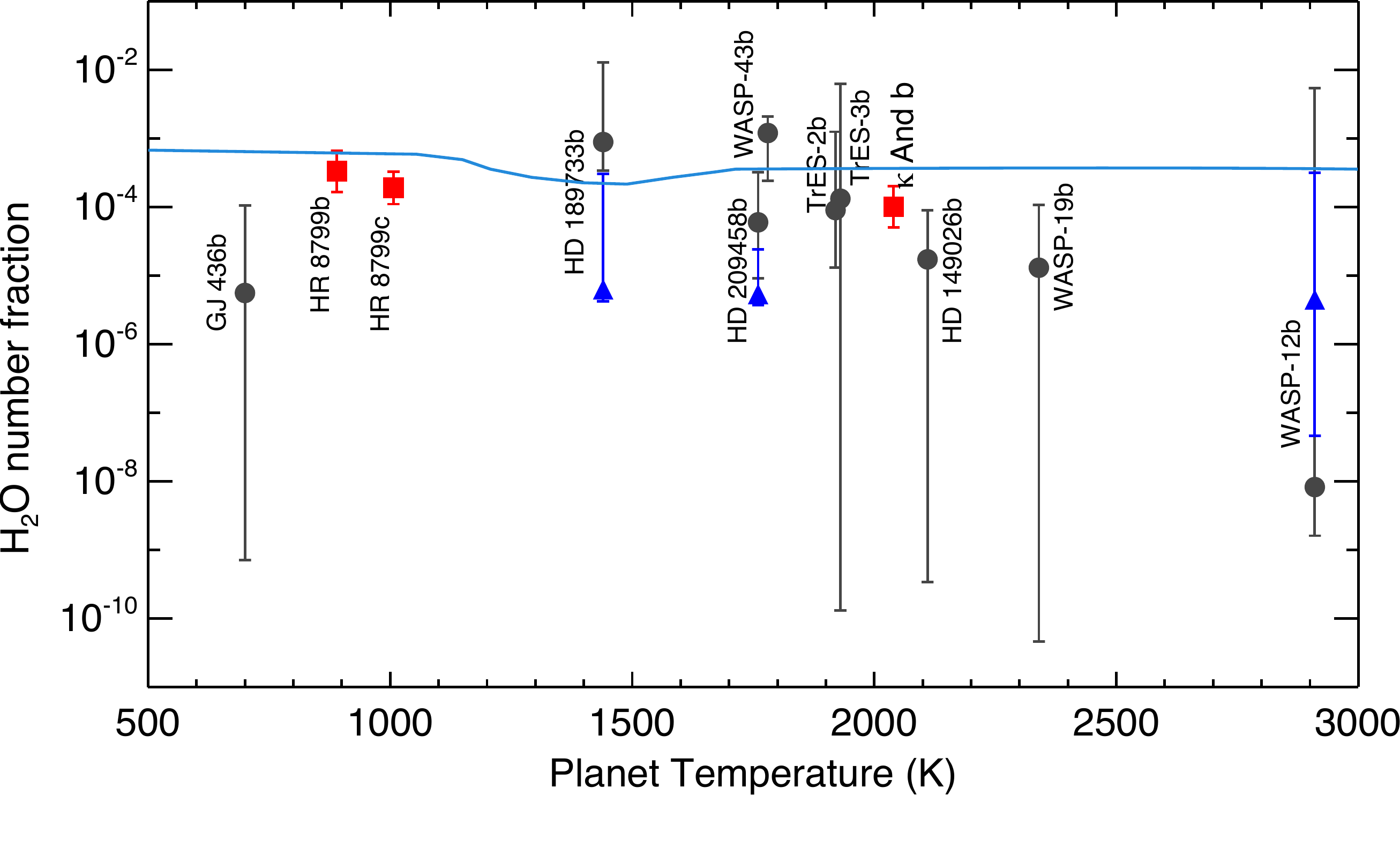}
\caption{\label{fig:water} \water\ abundances reported in extrasolar
  planets. Red squares indicate directly imaged planets, while the
  grey circles and blue triangles indicate transiting planet emission
  and transmission spectra, respectively. No clear trends with
  temperature (or planet mass) are apparent in the present data.
  Figure adapted from \cite{todorov:2015}, with recent data added
  \citep{diamondlowe:2014,barman:2015}. }
\end{center}
\end{figure}

\subsubsection{Spectroscopically Resolved Molecules}
\label{sec:molspec}
The most robust detections and measurements of extrasolar atmospheric
molecular abundances come from observations that resolve either the
individual lines, or (at lower dispersion) the overall shape of
molecular bandheads.

{\bf \hh} is of course the dominant constituent of most exoplanets
studied to date. Transiting planets with masses from Doppler
spectroscopy reveal planet densities that can only be fit with
significant \hh\ envelopes \citep{fortney:2007}. Substantial \hh\ can
also be inferred from the Rayleigh scattering slope of a planet's
transmission spectrum
\citep{lecavelier:2008haze189,lecavelier:2008haze209}. This feature
has been seen in both hot Jupiters and the $\sim$800~K Neptune-sized
GJ~3470b \citep{pont:2013,nascimbeni:2013,biddle:2014}, though a
recent analysis claims that starspots can masquerade as this Rayleigh
scattering signal \citep{mccullough:2014,oshagh:2014}.  \hh\ also
makes itself known via continuum-induced absorption (CIA), especially
in the $H$ and $K$ bands \citep{borysow:1997}. The CIA feature is
explicitly included when computing model spectra of all giant
extrasolar planets, so \hh's presence is implicit in these objects (as
it is in planet evolutionary models). However, no study has yet used
CIA to explicitly measure the atmospheric \hh\ mixing ratio.

{\bf CO} is abundant in giant, hot exoplanet atmospheres, and its
regularly-spaced rovibrational lines make it an especially easy
molecule to detect at high dispersion. In this way, the molecule has
been unambiguously detected in numerous hot Jupiters, though in most
cases the detection alone is not sufficient to tightly constrain CO's
atmospheric abundance
\citep{snellen:2010,brogi:2012,brogi:2014,rodler:2012,rodler:2013b,birkby:2013}.
CO has also been observed in several directly-imaged planets at medium
or low dispersion, including HR~8799b and~c
\citep{konopacky:2013,barman:2015} and beta~Pic~b
\citep{snellen:2014}. These studies generally reveal CO abundances
consistent with Solar metallicity and equilibrium chemistry, though
HR~8799b intriguingly shows only $\sim$50\% as much CO as predicted
\citep{barman:2015}. In all these planets, the CO/\methane\ ratio is
much higher than expected from equilibrium processes. This high CO
abundance is interpreted as evidence for strong quenching of these
species by vigorous vertical mixing (see Sec.~\ref{sec:diseq}.).

{\bf \water} is roughly as abundant as CO in the atmosphere (see
Fig.~\ref{fig:eq}), and indeed it sculpts their NIR spectra much more
strongly than does CO.  Again, medium- and high-dispersion
spectroscopy has detected the species in a rapidly-growing number of
directly imaging planets
\citep[e.g.,][]{patience:2010,gagne:2014,bonnefoy:2014a,barman:2015,todorov:2015}
and in several hot Jupiters
\citep{lockwood:2014,brogi:2014,birkby:2013}. Perhaps the largest body
of \water\ measurements comes from low-resolution spectroscopy of
transiting systems using HST's WFC3/G141 grism spectrograph, which
covers the 1.4\,\micron\ \water\ bandhead.  These observations reveal
\water\ absorption in many planets
\citep[e.g.,][]{deming:2013,nikolov:2015,kreidberg:2014b}, albeit
sometimes at low S/N.  Similar spectroscopy with Spitzer's IRS
spectrograph also hint at \water\ absorption at
6\,\micron\ \citep{grillmair:2008,todorov:2014}.  The lowest-mass
planet with observed \water\ is the $\sim$1000~K Neptune-sized
HAT-P-11b \citep{fraine:2014}. \water\ mixing ratios for all
characterized planets are plotted in Fig.~\ref{fig:water}: all
reported abundances are at least roughly consistent with equilibrium
conditions and a Solar-metallicity atmosphere \citep{todorov:2015}. As
with CO, \water\ abundance measurements of exoplanets should
considerably advance in the coming years: with spectra of directly
imaged planets from GPI and SPHERE, and by virtue of JWST's higher
sensitivity and broader wavelength coverage.

{\bf \methane} is expected to be less abundant than CO in all but the
coolest  exoplanets studied to date (see
Fig.~\ref{fig:eq}). Furthermore, even in cooler planets disequilibrium
processes tend to further decrease the \methane\ abundance in favor of
CO (see Sec.~\ref{sec:diseq}). The most convincing spectroscopic
detection of \methane\ to date is for the directly-imaged planet
HR~8799b, which exhibits a \methane\ abundance consistent with
expectations from a Solar-like metallicity and typical disequilibrium
conditions (see Fig.~\ref{fig:spectra}).  No conclusive,
spectroscopically resolved measurements of \methane\ have been made in
transiting planets, despite numerous attempts
\citep{swain:2008b,gibson:2011,dekok:2013,birkby:2013}. A peculiar
emission feature repeatedly and consistently detected from HD~189733b
at 3.3\,\micron\ has been attributed to
\methane\ \citep{swain:2010,waldmann:2012}, but this detection is
contested (see Sec.~\ref{sec:diseq}).  HST/WFC3 spectroscopy has the
sensitivity and wavelength coverage necessary to probe for
\methane\ in cool transiting planets (Benneke \&\ Morley, in prep), but no
detections have been made so far.

{\bf Other molecules:} At temperatures $\lesssim 500$~K, {\bf
  \ammonia} becomes increasingly abundant (see
Fig.~\ref{fig:eq}). This molecule has been detected in sub-equilibrium
abundances in cool brown dwarfs and solivagant (isolated, free-floating)
planetary-mass objects \citep{leggett:2013,line:2014c,line:2015}, but has not
yet been seen in any exoplanetary atmosphere.  At the warmer
temperatures of hot Jupiters, {\bf TiO} and VO could cause prominent
features similar to those seen in M stars \citep{reid:2005}. Tentative
spectroscopic evidence for TiO was reported in HD~209458b
\citep{desert:2008}, but subsequent high-dispersion spectroscopy shows
no evidence of the molecule \citep{hoeijmakers:2015}. Recent eclipse
observations of WASP-33b led to a claimed detection of TiO
\citep{haynes:2015}, but as the data show large systematics and
neither alternative optical absorbers nor reflection were considered
in the analysis, the detection should be regarded as tentative.

\subsubsection{Molecular Abundances from Broadband Photometry}
\label{sec:molphot}
Even when individual lines or bands are not resolved, a single
bandpass well-centered on a particular absorption or emission feature
can in principle constrain the abundance of the species of interest.
However, the data must be of high quality: while systematic noise is
unlikely to mimic the strengths and spacing of a high-dispersion CO
spectrum, correlated errors can more easily bias a single broadband
measurement.  Broadband measurements obtained at multiple epochs are
more susceptible to calibration drifts than is spectroscopy, for
example when comparing transit measurements when the host star is
variable and/or spotted
\citep[e.g.,][]{knutson:2011,pont:2013,barstow:2015}. Despite these
challenges, broadband data are almost always easier to acquire than is
spectroscopy, so photometry has historically preceded any dedicated
spectroscopic followup.

The Spitzer/IRAC camera  has observed photometric
transits and eclipses of more planets than any other facility, and has
often been employed in conjunction with ground-based imagers.  Though
early Spitzer photometry was used to claim various molecular
detections and atmospheric features (see Sec.~\ref{sec:inv}),
subsequent analyses significantly revised or cast doubt on many of the
past measurements
\citep{diamondlowe:2014,zellem:2014,lanotte:2014,evans:2015,morello:2015}
and led to the suggestion that broadband transit photometry may be
substantially less precise than previously claimed
\citep{hansen:2014spitzer}.  In light of these revisions it is debatable
whether broadband photometry usefully determines atmospheric
abundances in any transiting exoplanets.  Transit or eclipse
photometry with JWST may suffer from similar ambiguities, so plans to
use such techniques to, e.g., study habitable super-Earths'
atmospheres should be treated with great caution
\citep{kaltenegger:2009,deming:2009}.

Perhaps the most robust Spitzer result is the thermal emission
spectrum of GJ~436b, a transiting, $\sim$700~K, Neptune-mass
planet. Though reported secondary eclipse depths differ by
$>$4$\sigma$ in some bandpasses, analyses agree that planetary
emission is stronger in the 3.6\,\micron\ band but weaker at
4.5\,\micron\
\citep{stevenson:2010,stevenson:2012b,beaulieu:2011,lanotte:2014}.
Since the strongest absorbers are expected to be \methane\ at
3.6\,\micron\ and \coo\ and CO at 4.5\,\micron, numerous independent
analyses conclude that GJ~436b is enriched in CO and \coo\ and
depleted in \methane\ compared to equilibrium expectations for a
Solar-abundance atmosphere
\citep{madhusudhan:2011,line:2013,moses:2013}. The interpretation of
these analyses is discussed in Sec.~\ref{sec:diseq}.

Photometry seems somewhat more reliable for directly imaged
planets. The standard ground-based near-infrared bandpasses sample the
spectral energy distributions of a large number of these systems, but
to constrain atmospheric chemistry --- especially the relative
abundances of carbon-bearing species --- observations at longer
wavelengths ($\gtrsim$3\,\micron) are optimal. Ground-based photometry
at 3--5\,\micron\ has probed the atmospheres of several young,
directly imaged giant planets \citep[e.g.,][]{skemer:2014}, implying
CO/\methane\ ratios up to 100$\times$ greater than expected in
equilibrium (see Sec.~\ref{sec:diseq}). Broadband observations have
also led to inferences about cloud properties in these planets, as
discussed in Sec.~\ref{sec:clouds}.

Perhaps the most exciting ground-based photometric measurements to
date are of the 160~Myr-old GJ~504b \citep{kuzuhara:2013}. Photometry
on and off of the 1.6\,\micron\ \methane\ absorption band shows strong
emission outside the band and no detection within, indicating strong
\methane\ absorption \citep[][Skemer et al., in prep.]{janson:2013}
and perhaps the first robust detection of \methane\ in any exoplanet
atmosphere. With $T_{eff} \lesssim$600~K, GJ~504b is the coolest
exoplanet whose thermal emission spectrum has been studied. This
detection of \methane\ suggests that fundamentally new classes of
directly imaged planets are about to be studied; rapid progress will
likely be made in the coming years via followup studies of this and
other similarly cool planets with GPI, SPHERE, and JWST.

\subsection{Carbon-to-Oxygen Ratios}
\label{sec:c2o}
Carbon and Oxygen are the two most common elements in the Sun after H
and He \citep{asplund:2009}.  These two elements, so critical for life
on Earth, are expected to predominately form just a few dominant
molecular species in the planetary atmosphere studied to date when in
chemical equilibrium.  CO, \water, \coo, and \methane\ should all
induce prominent spectral features when present in sufficient
abundances.  A gas or ice giant's bulk composition --- and perhaps its
atmospheric makeup --- should be determined by the location in the
disk where the planet accretes most of its gas envelope
\citep{oberg:2011}.  In particular, the carbon to oxygen (C/O) ratio
of a planet's atmosphere will strongly affect the relative abundances
of these different molecules \citep[see Fig.~\ref{fig:eq}; also][Heng
\& Lyons, submitted]{madhusudhan:2012b,heng:2015c} and may hold clues
to the planet's formation and evolution (van Boekel et al., in prep.)

To date, there is at best tentative evidence for any exoplanet with a
C/O ratio significantly different from that of its host star and no
evidence for C/O$>$1.  Medium-resolution spectroscopy of the
directly-imaged planets HR~8977b and~c show their atmospheres to be
consistent with roughly solar C/O of $\sim 0.65\pm0.1$
\citep{konopacky:2013,barman:2015}, and high-dispersion spectroscopy
of the non-transiting hot Jupiter HR~179949b revealed a low-S/N
measurement of C/O$=0.5^{+0.6}_{-0.4}$ \citep{brogi:2014}.  These
spectroscopic measurements use data that directly resolve spectral
features from each particular C- or O-bearing molecule; they therefore
represent the current ``gold standard'' for these type of
measurements.

Observations of transiting gas giants, typically relying on broadband
photometry but with a growing number of spectroscopic data sets, also
find no unusual C/O ratios. A uniform analysis of nine
hot, giant transiting planets showed no evidence for C/O$>$1
\citep{line:2014}. A similar analysis of six additional hot Jupiters is
inconclusive: it showed that some broadband measurements were better
fit by models with C/O$>$1, but the final interpretation is unclear since no
uncertainties were reported for the derived planetary C/O values
\citep{madhusudhan:2012b}. A third, independent analysis of data from
six hot Jupiters again showed no evidence for C/O$>$1
\citep{benneke:2015}.

The most controversial C/O measurements have been for the hot Jupiter
WASP-12b. On the basis of broadband photometry, this bloated hot
Jupiter was the first planet claimed to have C/O$>$1
\citep{madhusudhan:2011}.  Additional eclipse measurements
\citep{cowan:2012,crossfield:2012d,swain:2013}, accounting for a
nearby M dwarf binary \citep{crossfield:2012d,bechter:2014}, and an
independent retrieval analysis \citep{line:2014} demonstrated that on
the contrary, existing data were insufficient to justify claims of
C/O$>$1.  New measurements and a reanalysis of old data, and
atmospheric retrieval including additional molecules (e.g., HCN)
generated a counter-claim that WASP-12b does show C/O$>$1
\citep[though this last analysis ignored multiple data points not
well-matched by the data;][]{stevenson:2014b}.  Most recently, new
HST/WFC3 data again point to a roughly Solar C/O ratio, leaving no
unusually carbon-rich planets in the field
\citep{kreidberg:2015,benneke:2015}.

High-S/N spectroscopy of directly-imaged planets seems the best
near-term hope for additional, more precise measurements of C/O
ratios.  GPI, SPHERE, and similar instruments should enable these
measurements for a growing sample of planets and measure their
atmospheric abundances. In the longer term, JWST spectroscopy of both
transiting and directly imaged planets will offer broader wavelength
coverage than Spitzer or HST and significantly improve C/O constraints
for transiting planets as well.


\subsection{Disequilibrium Chemistry and High Metallicity Atmospheres }
\label{sec:diseq}
Within the Solar System, no atmosphere is in chemical equilibrium and
all exhibit higher metallicity than the Sun. At high altitudes
photochemistry caused by the Sun's irradiation induces new reactions,
such as the formation of O$_3$ (ozone) from O$_2$ in the Earth's upper
atmosphere \citep{chapman:1930}.  Vigorous internal mixing deeper in
the atmosphere can ``quench'' abundances at higher altitudes, and is
responsible for the CO observed in Jupiter's cold atmosphere
\citep{prinn:1977}. As for metallicity, the Solar System's gas giants
become increasingly enriched in heavier elements with decreasing mass,
with Jupiter exhibiting $\sim$3$\times$ Solar abundances and the ice
giants $\gtrsim$50$\times$ \citep{atreya:2005,karkoschka:2011}; the
local terrestrial planets have negligible H$_2$ and so are
almost entirely `metallic' atmospheres.  

\begin{figure}[tb]
\includegraphics[width=3.5in]{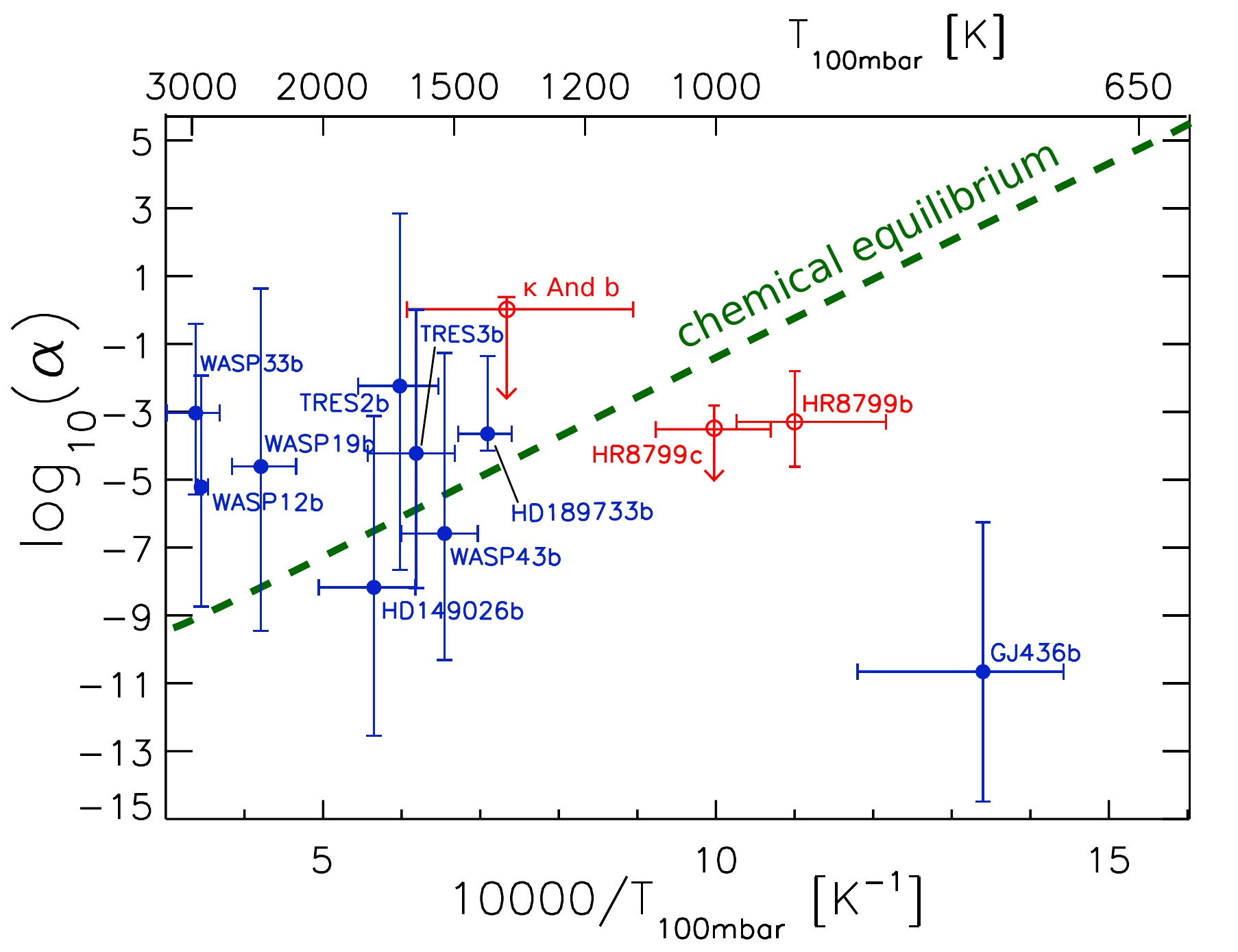}
\caption{\label{fig:diseq} Evidence for disequilibrium chemistry in
  exoplanetary atmospheres.  In equilibrium $\alpha$ should increase
  with decreasing $T$, as indicated by the dashed line.  Transiting
  plants are in blue, directly imaged planets are in red. Photospheric
  abundances in hotter planets are typically consistent with chemical
  equilibrium, but disequilibrium effects become increasingly apparent
  below $\sim$1200~K.  Adapted from \cite{line:2013b} and updated with
  recent data \citep{konopacky:2013,todorov:2015,barman:2015,haynes:2015}. }

\end{figure}

Given the important role that disequilibrium and elemental composition
play in shaping local planetary atmospheres, it is logical that they
should influence exoplanetary atmospheres as well. At the high
temperatures of the hottest hot Jupiters, kinetic reactions occur so
rapidly that all species are driven toward equilibrium abundances.
Disequilibrium processes should most strongly affect the atmospheres
of planets with $T\lesssim2000$~K
\citep[e.g.,][]{moses:2011,miguel:2014}, which includes all directly
imaged planets and most of the transiting planets.  \cite{line:2013b}
presents the modified equilibrium constant as a metric for assessing
the presence of disequilibrium conditions in a hot exoplanet's
atmosphere: $\alpha = \left( f_{CH_4} f_{H_2O} \right) / \left( f_{CO}
  f^3_{H_2} P^2 \right)$, where the $f_i$ are the mixing ratios of the
several species, $P$ is the atmospheric pressure level probed by
observations, and $\alpha$ is to be compared with the value expected
in equilibrium (see \citeauthor{heng:2015c} 2015 for an elegant
derivation of this relation and discussion of disequilibrium
conditions).  Current uncertainties on molecular abundances are large;
nonetheless Fig.~\ref{fig:diseq} shows $\alpha$ for exoplanet
atmospheres with useful abundance constraints. Cooler planets like
GJ~436b and HR~8799b may exhibit disequilibrium conditions, whereas
planets hotter than $\sim$1200~K are near equilibrium. The 3000~K
WASP-33b is a possible high-temperature outlier \citep{haynes:2015},
warranting further studies of this object.

\subsubsection{In Transiting Planets}
Most discussions of disequilibrium chemistry and high metallicity in
transiting exoplanet atmospheres relate to the thermal emission
spectrum of the $\sim$700~K hot Neptune GJ~436b, which is the current
best candidate for exhibiting strong disequilibrium effects
\citep{line:2013b}. Fig.~\ref{fig:eq} shows that in an equilibrium,
Solar-metallicity atmosphere at this temperature \methane\ should be
much less abundant than CO.  Yet as described in
Sec.~\ref{sec:molphot}, although the shape of the planet's spectrum is
still under debate multiple analyses agree that the planet emits at
3.6\,\micron\ (where \methane\ would absorb) but shows no emission at
4.5\,\micron\ (where CO and/or \coo\ would absorb).

While photochemistry alone cannot destroy enough \methane\ to explain
GJ~436b's spectrum, the combination of photochemical \methane\
destruction and vigorous internal eddy diffusion to quench
high-altitude \methane\ and CO could explain the observed spectrum
\citep{line:2011,madhusudhan:2011c,moses:2013}.  Alternatively, the
high CO/\methane\ ratio can be fit with an extremely high-metallicity
atmosphere, with an enhancement $\gtrsim$300$\times$ over Solar
\citep{moses:2013,fortney:2013}. Such an atmosphere might result if
only a modest H$_2$ envelope were accreted by the young planet; the
low H abundance naturally leads to a preference for H-poor molecules
like CO over H-rich molecules like \methane. An interesting twist on
the H$_2$-poor atmosphere hypothesis would be a He-dominated
atmosphere, left over after near-complete erosion of a primordial
H$_2$ envelope \citep{hu:2015}. The current data quality for GJ~436b
is too poor to discriminate between all these theories, but JWST eclipse
spectroscopy should demonstrate the extent to which disequilibrium,
high-metallicity, and unusual abundance patterns affect the atmosphere
of  GJ~436b and other sub-Jovian planets.

A possible but controversial indication of disequilibrium processes
comes from the hot Jupiter HD~189733b.  Ground-based spectroscopic
observations of this planet's thermal emission show unexpectedly
strong thermal emission at
3.3\,\micron\ \citep{swain:2010,waldmann:2012}, attributed to
\methane\ fluorescence analogous to that seen on Jupiter, Saturn, and
Titan \citep{drossart:1999,kim:2000}.  The feature could potentially
be produced by incorrect data calibration
\citep{mandell:2011,crossfield:2012a}, but the consistent detection of
this spectral feature from HD~189733b on multiple nights and the
absence of any such feature in otherwise identical observations of
HD~209458b \citep{zellem:2014b} suggests that the detection may be
genuine.  Eclipse spectroscopy of HD~189733b with JWST will easily
detect any such feature, determine whether it is caused by \methane,
and so settle the matter.

\subsubsection{In Directly Imaged Planets}
The higher-quality data available for directly imaged planets allows
more detailed studies of atmospheric chemistry and disequilibrium than
is possible for transiting systems.  Spectroscopy of the accessible
planets reveals that they, as for the transiting GJ~436b described
above, exhibit CO/\methane\ ratios significantly greater than would be
expected from equilibrium models and Solar abundances
\citep{barman:2011,barman:2011b,barman:2015,lee:2013,skemer:2014}. Stellar
irradiation and photochemistry is negligible for these young, hot
planets, so vigorous internal eddy diffusion is presumably responsible
for quenching the upper-atmospheric abundances at the equilibrium
values lower in the planets' atmospheres. Curiously, despite the
high-quality data available for directly imaged planets molecular
abundances for these systems are less frequently reported than for
transiting planets. With spectroscopy of many additional systems
expected soon from high-contrast spectrographs, the field would
benefit from a more systematic analysis of these planets' atmospheric
conditions and molecular abundances (see Sec.~\ref{sec:retrieval}).


\subsection{Alkalis, Ions, and Exospheres}
\begin{figure*}[tb]
\centering
\includegraphics[width=5.5in]{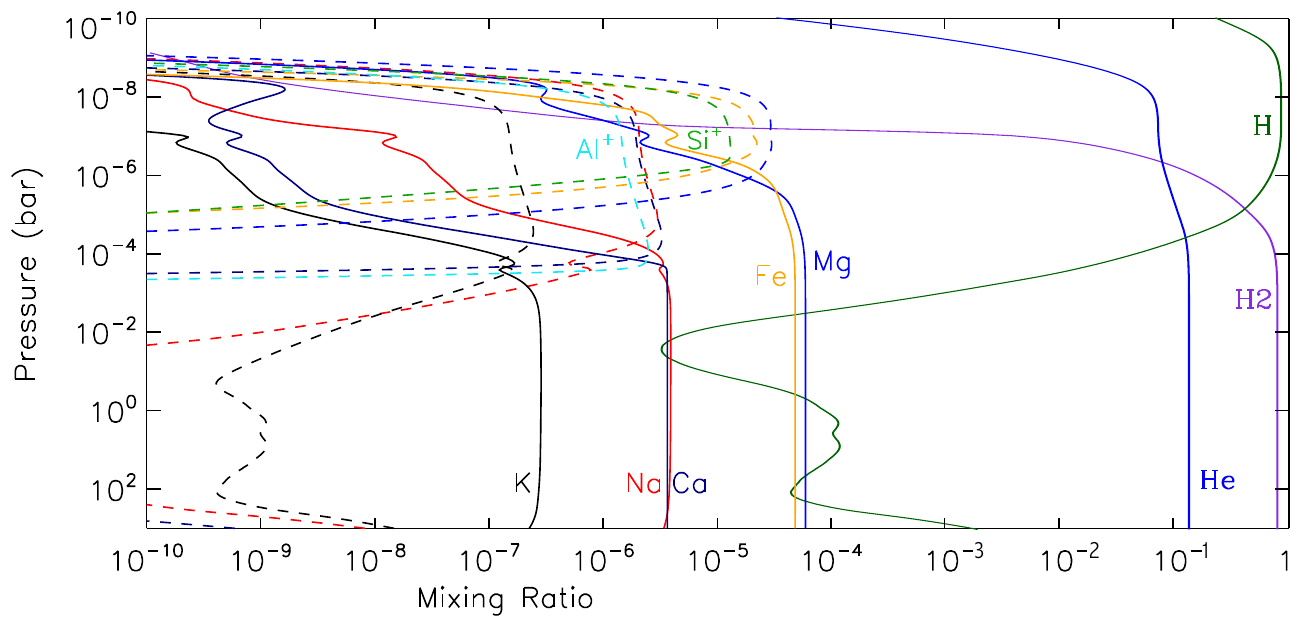}
\caption{\label{fig:exosphere} Theoretical abundance profiles of
  atomic and ionized species in the atmosphere of hot Jupiter
  HD~209458b. Solid lines are neutral species; dashed lines of the
  same color indicate the corresponding ion. Observations of these
  species probe mbar-to-$\mu$bar levels, much higher than the
  bar-to-mbar levels probed by thermal emission measurements.
  Adapted from \cite{lavvas:2014}. }
\end{figure*}

Highly irradiated planets such as hot Jupiters have temperatures on
their day sides up to roughly 3000~K --- as hot as some M stars --- at
pressures of $\sim$1~bar.  At lower pressures (higher altitudes), the
atmospheric density steadily decreases until the conditions for local
thermodynamic equilibrium are no longer met.  In the high-altitude
exosphere, temperatures can reach up to $\sim$10,000~K, hot enough to
split molecules and partially ionize the constituent
atoms. Fig.~\ref{fig:exosphere} shows theoretical abundance profiles
of some of the more common species predicted to exist in these
planets' atmospheres \citep{lavvas:2014}. In some cases the abundances
are predicted to be even greater than shown here as a planet's
\hh-dominated atmosphere escapes to space and heavier species are
entrained in the resulting outflow. The studies of these several phenomena are summarized in the recent review by \cite{fossati:2015}.

Alkali species such as sodium were the first constituents detected in
a hot Jupiter's atmosphere \citep{charbonneau:2002,redfield:2008}.
Mass loss and high-altitude atomic species have been observed in a
growing number of short-period transiting planets using ultraviolet
HST transit spectroscopy. These detections include increased
absorption in H$\alpha$ \citep[e.g.,][]{vidal-madjar:2003},
oxygen \citep{vidal-madjar:2004}, carbon and silicon
\citep{linsky:2010}, and tentative detections in
other high-order ultraviolet metal lines
\citep{fossati:2010,haswell:2012}.




Mass loss has been observed for a number of hot Jupiters through the
detection of extremely deep transits in excitation lines of hydrogen
and various metals, especially via ultraviolet transit spectroscopy
with HST
\citep{vidal-madjar:2003,vidal-madjar:2004,linsky:2010,fossati:2010}.
Qualitatively, the mass loss is understood to result from hydrodynamic
Roche lobe overflow of the planet's exosphere, powered by the
extremely high-energy X-ray and FUV flux of the host star.  However,
quantitative models of this phenomenon produce mass loss rates and
conditions in the exosphere which vary by several orders of magnitude
\citep{penz:2008,murray-clay:2009,koskinen:2012a} indicating that the
detailed mechanisms involved remain only poorly understood.

Atmospheric mass loss from lower-mass planets is even less well
studied than from hot Jupiters; the only observation to date are
Ly~$\alpha$ transit observations of hot Neptune GJ~436b and the
$\sim$2000~K sub-Neptune 55~Cnc~e. The nondetection of Ly~$\alpha$
around the latter sets an upper limit to the planet's hydrogen
mass-loss rate of $\sim 3\times 10^{8} \textrm{g~s}^{-1}$
\citep{ehrenreich:2012}.  More exciting is the substantial hydrogen
envelope recently detected around GJ~436b, seen as a $\sim$50\%
transit depth in Ly~$\alpha$ and indicating a mass-loss rate of
$\sim10^{8}$--$10^{9} \textrm{g~s}^{-1}$ \citep{ehrenreich:2015}. The
upper range of both these measurements are a factor of a few lower
than those predicted by \cite{lammer:2013}, and 10--100 times lower
loss rates predicted by other studies \citep{valencia:2010,owen:2012}.


\section{Clouds \& Hazes}
\label{sec:clouds}
\begin{figure}[tb]
\begin{center}
\includegraphics[width=3.2in]{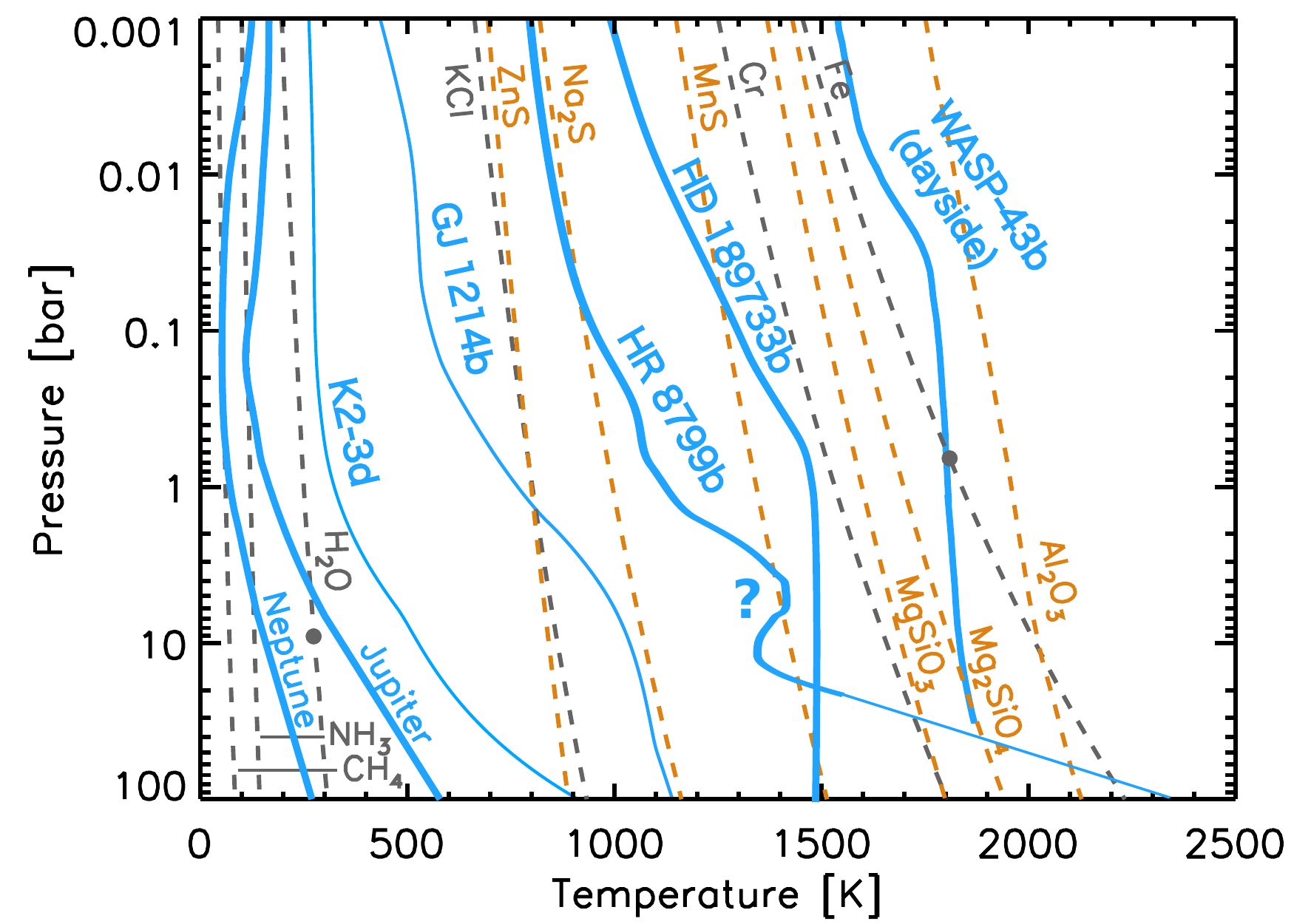}
\caption{\label{fig:condense} Condensation curves from
  \cite{marley:2014b} for various refractory species assuming Solar
  metallicity (dashed lines) compared to temperature-pressure profiles
  of known planets (solid lines). Thick lines are empirical
  measurements from atmospheric retrieval analyses
  \citep{lee:2013,crouzet:2014,stevenson:2014c}; thin lines are
  theoretical models \citep[][T.~Barman et al., private
  communication]{crossfield:2011}.  The retrieved exoplanet profiles
  are all essentially unconstrained at $\gtrsim$3~bar and $\lesssim
  10$~mbar. Thus the kink in HR~8799b's profile could be either a
  numerical artifact or the result of a thick, low-altitude cloud
  layer.  }
\end{center}
\end{figure}

As with disequilibrium chemistry (see Sec.~\ref{sec:diseq}), clouds
and hazes are ubiquitous in all Solar system planets with thick
atmospheres.  A growing body of evidence indicates that atmospheric
condensates are also common in extrasolar planet atmospheres. Clouds
form when gaseous species condense out of the atmosphere, forming
liquid droplets or solid `dust' particles.  Fig.~\ref{fig:condense}
shows condensation curves for compounds that may be present in
exoplanetary atmospheres. These clouds may form at points where the
condensation curves intersect a planet's atmospheric thermal
profile. It is clear from Fig.~\ref{fig:condense} that a wide
diversity of clouds are possible. The section below first discusses
the early signs of clouds in brown dwarfs and more recent signs in
their less-massive, directly imaged cousins, and follows with the
evidence for clouds and haze in transiting exoplanets.  Interested
readers are strongly recommended to read the in-depth review by
\cite{marley:2013}, which gives a much more detailed description of
clouds and hazes in both exoplanets and brown dwarfs.

\subsection{Brown Dwarfs and Young, Massive Planets}
\label{sec:bdclouds}
As brown dwarfs age and slowly cool, clouds of condensed refractory
compounds (e.g., calcium aluminates, silicates, and iron) start to
form in their atmosphere \citep{lunine:1986,lodders:1999}. As the
cooling continues, brown dwarfs transition from dusty atmospheres
\citep{kirkpatrick:1999,martin:1999} to dust-free conditions
\citep{burgasser:2001,burgasser:2002} when the effective temperature
reaches $\sim$1300 K \citep{stephens:2009,testi:2009}. Different
models have been proposed to explain this dust dispersal and include
either a sudden clearing or collapse of the cloud
\citep{stephens:2009,allard:2003,knapp:2004, tsuji:2003,tsuji:2004} or
patchy regions of varying cloud thickness
\citep{burgasser:2002,ackerman:2001,marley:2010} as seen on Jupiter
and Saturn.  To the extent that brown dwarfs are just more massive
versions of young, directly imaged planets, the same trends should
hold. However, the lower gravity of planets means that clouds remain
in the visible photosphere for much fainter planets than for brown
dwarfs \citep[e.g.,][]{barman:2011}.  Broadband photometry of many
directly imaged planets confirms that clouds are required to match
atmospheric models to observations, even when the planet is
substantially cooler than cloud-free brown dwarfs
\citep{madhusudhan:2011d,currie:2011,skemer:2014}. The near-infrared colors of
the cooler, fainter GJ~504b suggests that the transition to more
nearly cloud-free conditions occurs at lower temperatures than in
brown dwarfs \citep{kuzuhara:2013}.

Y dwarfs are objects with $T_{eff} \lesssim$450~K and low masses
($5-30 M_{Jup}$) that approach or straddle the canonical
planetary-mass boundary
\citep{cushing:2011,kirkpatrick:2012,leggett:2013,dupuy:2013,beichman:2014}.
In addition to substantial ammonia (NH$_3$), their atmospheres contain
many types of cloud species.  In the warmest Y dwarfs these clouds
should be composed of Na$_2$S, and indeed warmer Y dwarfs are better
fit by atmospheric models including clouds than by cloud-free models
\citep{morley:2012,beichman:2014}.  At temperature $\lesssim 300$~K
the condensed species may include H$_2$O, NH$_3$, and other, more
exotic species \citep{burrows:2003,visscher:2006,morley:2014}. No
models match the existing broadband data for the coolest objects
\citep[$\lesssim 250$~K;][]{luhman:2014,kopytova:2014}, emphasizing
that our understanding of these cool, cloudy planetary-mass objects
are still quite uncertain.

\subsection{Transiting Planets}
From the earliest models of transiting exoplanet atmospheres
\citep{seager:2000,sudarsky:2000,brown:2001,hubbard:2001} it was clear
that clouds should have a major impact on these planets' observable
properties.  The effect on transmission measurements can be especially
severe. The transit geometry increases the optical depth of absorbers
$\gtrsim50\times$ over that in the normal, thermal emission geometry,
which means that even optically thin species can obscure the
atmospheric composition \citep{fortney:2005}.

The most robust evidence of condensate particles in a transiting
exoplanet's atmosphere comes from spectroscopy of the $\sim$600~K,
2.6~\Rearth\ sub-Neptune GJ~1214b. A steady stream of transit
spectroscopy, capped by the high-precision HST/WFC3 spectroscopy shown
in Fig.~\ref{fig:spectra}, demonstrates that GJ~1214b has an
exquisitely flat transmission spectrum
\citep{bean:2010,bean:2011,crossfield:2011,berta:2012,demooij:2012,kreidberg:2014}. The
measurements rule out cloud-free atmospheres with absorption from most
spectroscopically active species (see Fig.~\ref{fig:spectra}), so a
high-altitude cloud deck at $P\lesssim0.1$~mbar seems the most likely
interpretation.  The aerosol's composition is unknown, but theories
include either condensate clouds of ZnS or KCl \citep[see
  Fig.~\ref{fig:condense};][]{morley:2012} or photochemically-produced
hydrocarbon hazes \citep{kempton:2012}.  A haze at such high altitude
obscures all but the strongest absorption features;
\cite{barstow:2013} suggest that transit spectroscopy with JWST may be
the best hope to descry features from either \methane\ (at
3.3\,\micron\ and 7.5\,\micron) and/or \coo (at 4.3\,\micron\ and
15\,\micron).

Several other sub-Jovian transiting planets also show fairly flat
transmission spectra, but in these cases clouds are required only if
the planets have H$_2$-dominated, low-metallicity atmospheres
\citep[cf.\ Sec.~\ref{sec:diseq};][]{knutson:2014a,knutson:2014b,fraine:2014,biddle:2014,ehrenreich:2014}.
Along with GJ~1214b, all these targets are small ($R<5$\Rearth) and
cool ($T_{eq} \lesssim 900$~K), which may provide a clue to the nature
of the obscuring material \citep{howe:2012,kempton:2012a,morley:2013}.

Larger, usually hotter transiting planets appear to be qualitatively
different. Most show \water\ absorption in HST/WFC3 spectroscopy,
albeit weaker than expected \citep[for a recent summary
  see][]{benneke:2015}. The interpretation seems to be that some hot
Jupiters have a thin cloud or haze layer that weakens (but does not
entirely mask) the signature of \water\ absorption
\citep[e.g.,][]{mccullough:2014,nikolov:2015,sing:2015}. As with the
smaller, cooler planets the cloud species and size distribution
remains unknown.  JWST spectroscopy offers a promising avenue for
determining the composition of these planets' clouds. Sub-micron
silicate cloud particles could create strong spectral features in
these planets' spectra at $\sim$10\,\micron\ \citep{lee:2014};
nonsilicate materials (e.g., tholins) may show similarly strong
features at other mid-infrared wavelengths \citep{wakeford:2015}.

Clouds, hazes, or dust have also been suggested for transiting
planets showing no conclusive features of any kind in emission
\citep{hansen:2014spitzer,evans:2015}, or showing only flat or sloped
transmission spectra at optical and infrared wavelengths
\citep{lecavelier:2008haze209,sing:2009haze,gibson:2013,sing:2013,pont:2013}. These
slopes have been interpreted as the signature of particulate Rayleigh
scattering in the upper atmosphere, which conveniently and plausibly
accounts for the muted spectral features discussed above and (by
absorbing starlight at altitude and so shifting the thermal profile)
for the near-blackbody emission spectra of some planets
\citep{crossfield:2012d,knutson:2012,pont:2013,burrows:2014}.

However, the interpretation  of Rayleigh scattering has been questioned in
 analyses of HD~189733b and other planets, with the slope
instead attributed to a haze-free atmosphere and transit contamination
from a nonuniform stellar surface
\citep{mccullough:2014,oshagh:2014}. High-precision optical albedo
measurements could settle the matter: by constraining the
presence and size distribution of any haze particles in these
atmospheres \citep{evans:2013,barstow:2014}. If the particles are
$\lesssim$1\,\micron, atmospheric circulation models indicate that
they should be present throughout the planet's atmosphere
\citep{parmentier:2013} -- and in particular, near the terminator
probed by transmission spectroscopy.

\section{Atmospheric Circulation \& Energy Budgets}
\label{sec:energy}
Planetary atmospheres typically exhibit variability and weather
patterns on many scales, from terrestrial hurricanes to Jupiter's
Great Red Spot to Saturn's dramatic storm of 2010
\citep{sromovsky:2013,sayanagi:2013}. The atmospheric processes that
drive all this global weather \citep[for a recent review
  see][]{heng:2015} must also play a role in the atmospheres of
exoplanets, and so the observed atmospheric properties of extrasolar
planets must also be variable at some level.  Another wrinkle for
short-period planets such as hot Jupiters is that these planets'
intense irradiation also leads to severe longitudinal temperature
gradients, observed as thermal phase curves.  The irradiation drives
the planets' circulation and the net input energy is necessarily
modulated by the planetary albedo.  All these processes combine to
determine a planet's thermal structure.  The section below discusses
each of these phenomena in turn: phase curves in
Sec.~\ref{sec:phaseobs}, albedos in Sec.~\ref{sec:albedo}, rotation
and variability in Sec.~\ref{sec:weather}, and thermal structure (and
the lack of any definitive thermal inversions) in Sec.~\ref{sec:inv}.

\subsection{Phase curves}
\label{sec:phaseobs}

\begin{figure*}[tb]
\includegraphics[width=3.1in]{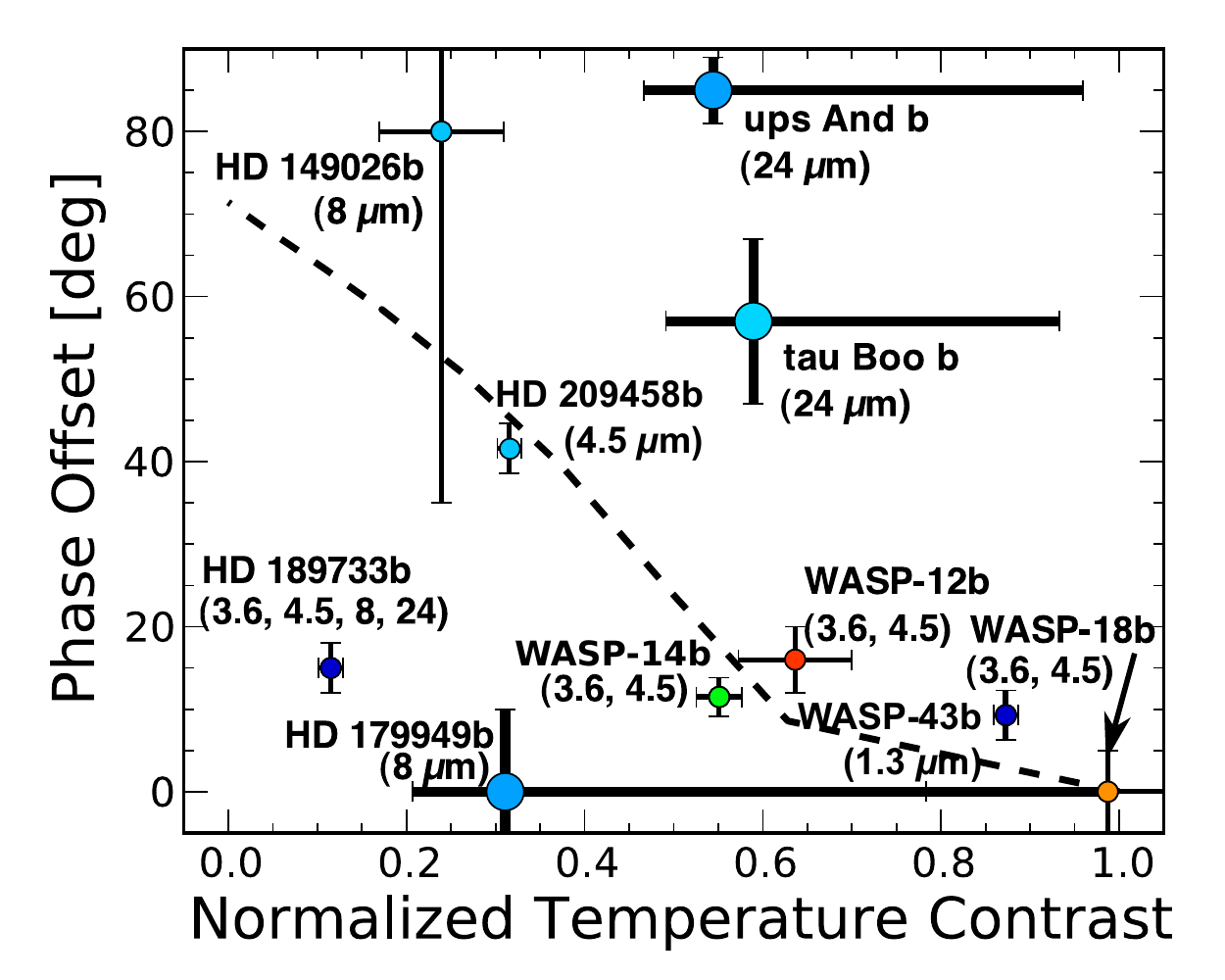}
\includegraphics[width=3.7in]{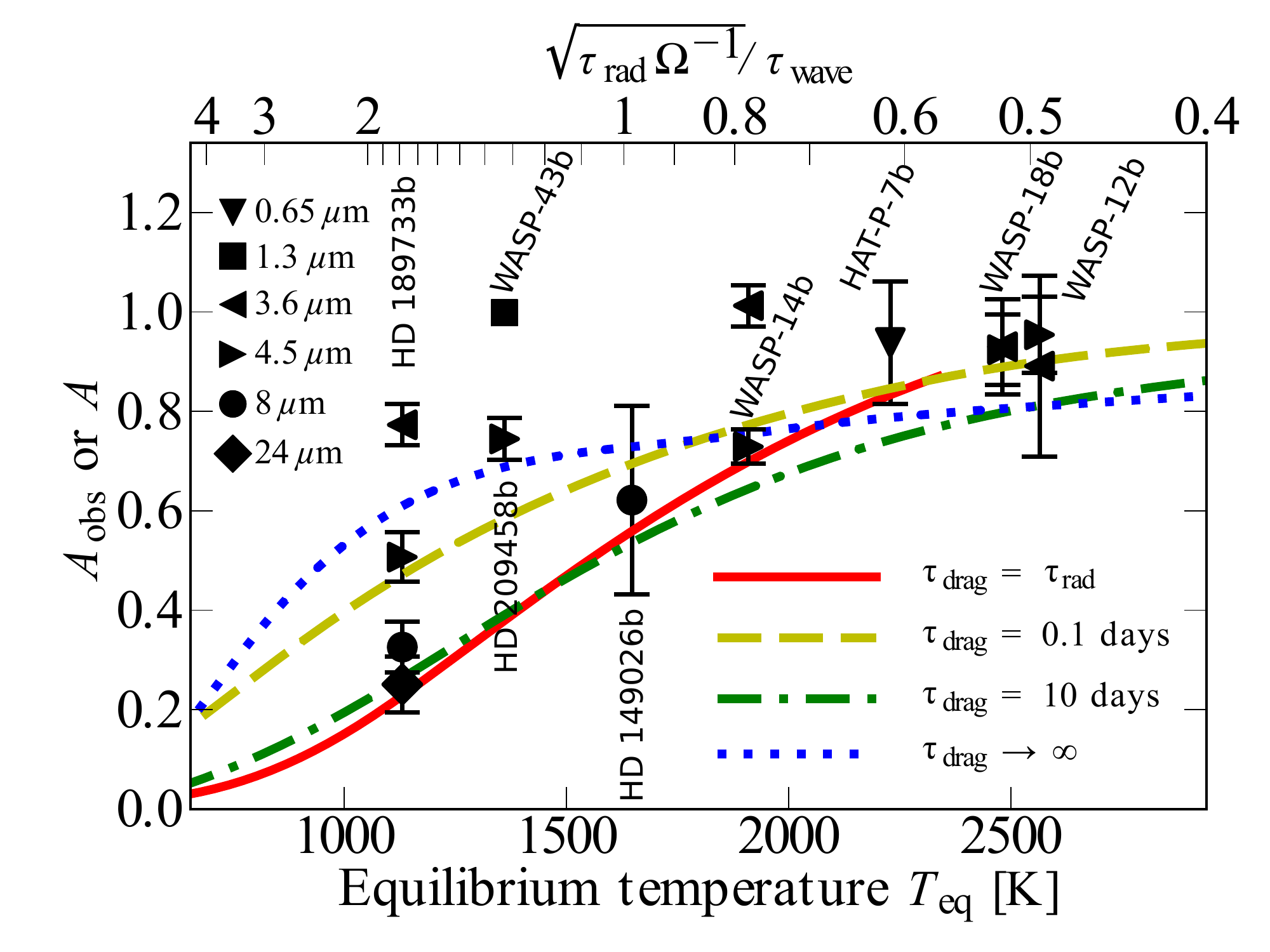}
\vspace{-0.1in}\caption{: \footnotesize {\em Left:} All infrared phase
  curve observations for hot Jupiters on circular orbits.  Temperature
  contrasts have been normalized by $T_\textrm{eq}$ (indicated by blue
  and red colors for cooler and hotter planets, respectively).  The
  dashed line is the semi-analytic radiative/advective model of
  \cite{cowan:2011circ}, which predicts smaller phase offsets for
  larger day/night temperature contrasts.  {\em Right:} Relative phase
  curve amplitudes vs.\ equilibrium temperature, along with
  shallow-water circulation model predictions for various drag
  timescales. Figure adapted from \cite{perezbecker:2013}, with
  additional data added \citep{zellem:2014,stevenson:2014c,wong:2015}.
  \label{fig:scatter}}
\end{figure*}

To date, phase curves have been measured at different wavelengths,
with different cadences, and with varying orbital coverage fractions.
Early phase curves of non-transiting hot Jupiters were of limited
utility because the then-unknown orbital inclination of our targets
introduced degeneracies in the interpretation of these data
\citep{cowan:2007,crossfield:2010}.  Now that high-dispersion
spectroscopy breaks the $\sin i$ degeneracy (see Sec.~\ref{sec:hds}),
non-transiting phase curves become more useful. Nonetheless most
observations to date focus on transiting systems.

One of the great successes of modeling exoplanet atmospheres was the
prediction of super-rotating equatorial winds on hot Jupiters, which
leads to hot spots shifted eastward from the substellar point
\citep{showman:2002}. This phenomenon is commonly observed in hot
Jupiter phase curves, with eastward phase offsets observed for dozens
of systems (see Fig.~\ref{fig:scatter}).  However, predictions that
day/night temperature contrast (and thus phase curve amplitude) should
increase with dayside temperature have been disappointed. Some
extremely hot Jupiters behave as expected (WASP-12b, WASP-18b), but as
shown in Fig.~\ref{fig:scatter} other, much cooler planets also
exhibit strong temperature contrasts (e.g., WASP-43b). It is sobering
to consider the utility of the data plotted in Fig.~\ref{fig:scatter}
if these Spitzer phase curve observations were found to suffer from
the same limitations and systematics seen in Spitzer broadband
photometry (Sec.~\ref{sec:molphot}).

Current phase curve observations cannot be explained by advection
alone, which generically predicts small phase offsets for
planets with large temperature contrasts
\citep{fortney:2008,cowan:2011circ}.  Fig.~\ref{fig:scatter} shows
that this trend is emphatically not observed, which could indicate any
combination of dynamically evolving atmospheric chemistry
\citep{agundez:2012}, inhomogeneous clouds \citep{showman:2013}, or
atmospheric drag forces \citep{rauscher:2013,perezbecker:2013}.  

Though most phase curves to date have been obtained in one bandpass at
a time, recent observations of hot Jupiter WASP-43b reveal the power
of phase curve spectroscopy \citep{stevenson:2014c}. This analysis
determined the planet's mean \water\ abundance and thermal structure
as a function of longitude, finding a much colder nightside than was
expected. Such spectroscopy will likely be used to great effect in the
coming years with both HST and JWST.

Phase curves have also been observed from many planets at optical
wavelengths, where scattered starlight is at least as important as
thermal emission. These analyses often show westward-shifted phase
offsets --- the reverse of what infrared studies have found. The
interpretation has been that reflective clouds form in the planets'
cooler regions, reducing thermal emission and boosting reflection
\citep{demory:2013b,shporer:2015,webber:2015,esteves:2015}. While this hypothesis is
plausible, caution is advised because the optical modulations could be
caused by either scattering in the planet's atmosphere or by
star-planet interactions.  Optical photometry of the hot Jupiter
$\tau$~Boo~b revealed optical modulation consistent with a planetary
phase curve, but with an amplitude much too large to be of planetary
origin \citep{walker:2008}.  It seems possible that the more recent
detections of optical modulation could be caused by a weaker form of
the $\tau$~Boo effect instead of having a solely planetary nature.

\subsection{Albedos}
\label{sec:albedo}
\begin{figure}[tb]
\begin{center}
\includegraphics[width=3.3in]{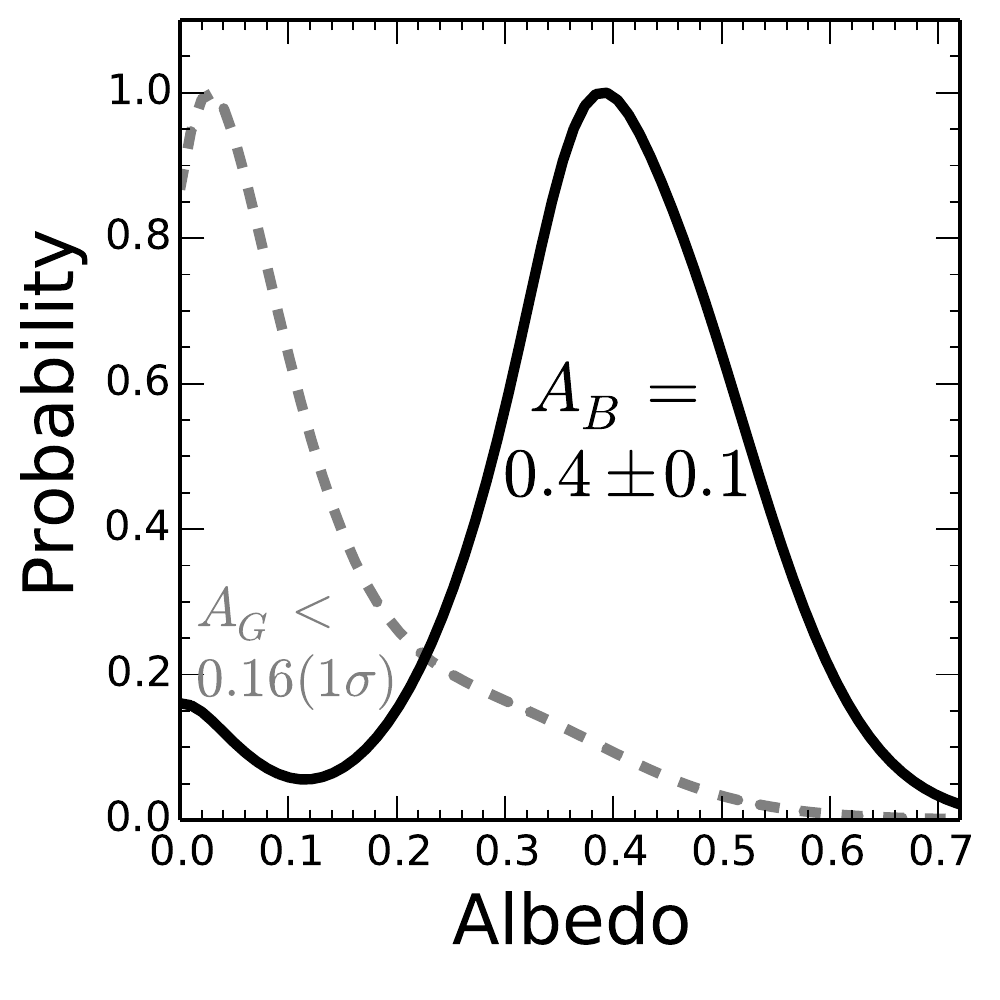}
\caption{\label{fig:schwartz} Constraints on hot Jupiter albedo and
  atmospheric recirculation, demonstrating the curious dichotomy
  between Bond and geometric albedos. Solid lines indicate Bond
  albedos of planets with infrared flux measurements; these data
  indicate $A_B \approx 0.4$. Dotted outlines indicate geometric
  albedos of planets with optical secondary eclipses, typically with
  Kepler; these data indicate $A_G \lesssim 0.16$ in the Kepler
  bandpass. To explain this striking discrepancy, hot Jupiters must
  have $A_G \gtrsim 0.5$ beyond the Kepler bandpass, at $\lambda <
  0.4$\,\micron\ and/or $\lambda > 0.8$\,\micron.  Data taken from
  \cite{schwartz:2015} and \cite{wong:2015}.  }
\end{center}
\end{figure}

Broadband photometry of hot Jupiters indicates that they typically
have low geometric albedos at optical wavelengths
\citep[$A_G\lesssim0.1$; e.g.,][]{rowe:2008}. In contrast, thermal
emission measurements (mostly from Spitzer secondary eclipses)
indicate that hot Jupiters' Bond albedos are typically $A_B \approx
0.4$ \citep{schwartz:2015}.  Fig.~\ref{fig:schwartz} demonstrates this
curious dichotomy.  Most planets studied in this way have been
observed with either Kepler alone (and orbit stars too faint for
followup with Spitzer or HST) or only with Spitzer (and do not lie in
the Kepler field of view). Nonetheless these discrepent families of
albedo measurements do not arise from {selection effects}: HD~189733b
and HD~209458b have been observed at both optical and NIR wavelengths,
and both show the same discrepancy with $A_B>A_G$.  Similarly, no
obvious biases result from the use of Spitzer data since consistent
Bond albedos are reported for WASP-43 whether only HST or both HST and
Spitzer data are used \citep{stevenson:2014c,schwartz:2015}. The
planets exemplifying this albedo phenomenon span a wide range of
temperatures and are unlikely to host the same cloud species, so this
issue transcends any single class of planets.

The most likely solution to this issue is that hot Jupiters
have substantially higher geometric albedos beyond the optical
bandpasses used \citep{schwartz:2015}. Indeed, the first extrasolar
albedo spectrum \citep[of HD~189733b;][]{evans:2013} revealed $A_G=0$
within the Kepler bandpass but $A_G=0.4$ at 300--450~nm.  These
observations can be fit by a wide range of atmospheric compositions
including any of alkalis, clouds/hazes, or even gaseous TiO/VO
\citep{barstow:2014}. Such models would also naturally explain the low
$A_G$ of HD~209458b in the MOST bandpass \citep{rowe:2008}. However,
if the estimate of $A_B\approx0.4$ is correct
(cf.\ Fig.~\ref{fig:schwartz}) then hot Jupiters may also reflect
considerable starlight at $\gtrsim$800~nm. Optical eclipse spectroscopy
can best test these theories by directly measuring hot Jupiters'
geometric albedos as a function of wavelength.

\subsection{Rotation, Variability, \& Weather}
\label{sec:weather}

Despite the importance of rotation rate, internal heat flux, and
external irradiation on planetary weather, these factors are not
typically known for extrasolar planets. Short-period giant planets
such as hot Jupiters are thought to be tidally locked
\citep{correia:2010}. In principle phase curve observations can
constrain a planet's rotation rate and test for tidal locking, but the
dependence is highly degenerate with other parameters
\citep[e.g.,][]{showman:2009}. Lower-mass, short-period planets will
not necessarily be tidally locked, which will render any
interpretation of these planets' phase curves (e.g., with JWST) even
more uncertain.  High-dispersion spectroscopy can also constrain
planetary rotation \citep{kempton:2012,kempton:2014,rauscher:2014},
but to date the only observation is a tantalizing hint of windspeeds
measured at the terminators of HD~209458b and HD~189733b
\citep{snellen:2010,wyttenbach:2015}.

The best rotation measurement to date came from high-dispersion
spectroscopy of the directly imaged planet beta~Pic~b, which shows a
projected equatorial velocity of 25~km~s$^{-1}$
\citep{snellen:2014}. Assuming a radius of $\sim$1.5\,$R_J$
\citep{bonnefoy:2014b}, this rotation period of $\sim$7~hr is shorter
than that of Jupiter or Saturn and also less than most brown dwarfs or
low-mass stars at a comparable age \citep{bouvier:2014}.  The rotation
will presumably increase further as the planet ages, cools, and
contracts.  High-dispersion spectrographs on the next generation of
giant telescopes will measure $v \sin i$ for many additional planets,
and also create global 2D weather maps via Doppler imaging
\citep{crossfield:2014a,crossfield:2014b}.

For giant planets that rotate at least as quickly as beta~Pic~b,
detection of rotationally-modulated variability should be feasible in
the future.  Observations of brown dwarfs reveal rotation rates and
variability for many objects
\citep[e.g.,][]{artigau:2009,radigan:2014}. Variability seems
ubiquitous for brown dwarfs of all spectral types
\citep{buenzli:2014,metchev:2015}, consistent with rotational
modulation of an inhomogeneous, cloudy surface as discussed in
Sec.~\ref{sec:bdclouds} above.  In the case of brown dwarfs, the
variability is variously attributed to inhomogeneities in clouds
\citep{apai:2013} or internal thermal perturbations
\citep{robinson:2014}. High-contrast photometry of directly imaged
planets should constrain these same processes in planets rather than
brown dwarfs \citep{kostov:2013}.

Some early general circulation models of hot Jupiters predicted that even
these highly-irradiated planets should not be in a perfectly steady
state, but should show substantial inter-epoch variability in thermal
emission \citep{rauscher:2007a}. Only a handful of these planets have
been observed multiple times with the precision necessary to detect
variability. The tightest limits come from seven 8\,\micron\ eclipses
of HD~189733b, which ruled out variability above 2.7\%
\citep{agol:2010}. Recent Spitzer observations of 55~Cnc~e's eclipses
suggest that the planet's thermal emission may vary with time, but the
underlying cause remains unknown \citep{demory:2015}. Variations in
transit depth are more difficult to interpret, since they can much
more easily result from starspots and stellar variability rather than
from the planet \citep{agol:2010,knutson:2011}. JWST's enhanced
sensitivity ($\sim$8$\times$ greater than Spitzer's) should allow it
to place much tighter constraints on variability of thermal emission,
even if transit variability remains dominated by stellar effects.

\subsection{Looking for Thermal inversions}
\label{sec:inv}
A planet's thermal structure is said to be inverted when at some
altitude its atmosphere satisfies $dT/dP<0$, i.e.\ when temperature
increases with increasing altitude.  All of the Solar system's outer,
giant planets exhibit temperature inversions above a pressure level of
roughly 0.1~bar \citep{orton:1981}, where the outgoing infrared
opacity drops and incident radiation begins to dominate
\citep{robinson:2013}.  Such inversions are typically characterized by
looking at the planets' thermal spectra, which shows small, narrow
emission features in the broader, deeper cores of absorption lines.
Isolated brown dwarfs and planetary-mass objects show no such
inversions \citep{lunine:1986,stephens:2009,line:2014c,line:2015}, but this
phenomenon has been long-sought in planets orbiting other stars.

It was appreciated soon after the discovery of hot Jupiters that their
intense insolation levels would lead to much shallower T/P profiles
\citep{seager:1998, goukenleuque:2000}, though it took several more
years to realize that short-wavelength absorbers with sufficient
optical depth might lead to a strong inversion in these atmospheres
\citep{hubeny:2003}. The claimed detection of just such an inversion
at $\sim$1~bar in HD~209458b seemed to vindicate these theories
\citep[][but see below]{knutson:2008} and was used to predict two
distinct classes of hot Jupiter atmospheres: those without inversions
and those with thermal inversions induced by some high-altitude
absorber \citep[][]{burrows:2008,fortney:2008}. New observations and
analyses of other targets seemed to confirm the inversion dichotomy
\citep[see][]{madhusudhan:2010,knutson:2010}.

However, it now appears that inversions as initially reported are not
present in these planets. In an early retrieval analysis of four
planets with claimed inversions, data for two planets were equally
well fit by models with or without temperature inversions
\citep{madhusudhan:2010}.  A second, similar analysis of nine planets
found that all but HD~209458b were best explained by non-inverted
atmospheres \citep{line:2013}. A later  analysis of new and archival
data of HD~209458b then showed that this prototypical inverted
atmosphere is not inverted after all
\citep{diamondlowe:2014,zellem:2014}. The best remaining candidates
for an inverted atmosphere are HAT-P-7b \citep{christiansen:2010},
which has not been subjected to any reanalysis or atmospheric
retrieval; and WASP-33b \citep{haynes:2015}. Even if these last two
planets do indeed host thermal inversions, it is clear that the
two-class model as originally proposed is now defunct.





\section{The Future}
\label{sec:future}
As much as any other advances, the considerable recent progress made
in characterizing exoplanet atmospheres has come from the development
of reliable spectrographic instruments and techniques and from the
continued discovery of of new planetary systems that can be more
easily observed at high S/N.  Progress in both these areas will only
accelerate in the coming years: spectroscopy with new facilities such
as JWST, WFIRST-AFTA, and ground-based extremely large telescopes, and
discoveries of many new planets from ongoing or imminent surveys from
GPI, SPHERE, TESS, and GAIA.

\subsection{The Near Term: 2015--2025}
Direct imaging surveys using GPI, SPHERE, and similar instruments are
already underway. These will discover and obtain low-resolution
spectroscopy of perhaps dozens of young, self-luminous planets. The
improved contrast provided by these facilities over existing
instruments (see Fig.~\ref{fig:contrast}) will also permit
higher-quality spectra of planets currently known. The astrometric
GAIA mission, also underway, is projected to find tens of thousands of
planets on several-AU orbits; a fraction of these will also be
accessible to characterization via direct imaging
\citep{perryman:2014,sozzetti:2014}. Spectroscopy of these systems
with ground-based systems and with JWST's significantly broader
wavelength coverage \citep{beichman:2010} will place ever-tighter
constraints on the composition, chemistry, and thermal structure of
giant planetary atmospheres.  Planetary variability and rotation rates
will provide insight into circulation and cloud processes for a
growing number of planets. Optical-wavelength observations may reveal
signs of ongoing accretion and planet formation (Sallum et al., in
prep.), and heroic efforts for a small number of systems could begin
to detect starlight reflected from old, cold gas giants.  Studies of
brown dwarfs will continue to lead the way in demonstrating what can
be learned about cool, substellar atmospheres from reliable,
high-quality data \citep[e.g.,][]{line:2015}.

The upcoming Transiting Exoplanet Survey Satellite (TESS) mission,
scheduled for launch in 2017, will find thousands of new terrestrial
planets, sub-Jovians, and hot Jupiters transiting bright stars
\citep{ricker:2014}. Many of these will be suitable for atmospheric
study with JWST via transits, eclipses, and phase curves
\citep{batalha:2013,beichman:2014}.  Proposals for JWST's Guaranteed
Time and Cycle~1 observations are scheduled for 2017, which means that
TESS' planets will arrive too late for observation in the first 1--2
years of JWST operations.  Fortunately, the K2 mission
\citep{howell:2014} is already finding planets suitable for JWST
transmission spectroscopy
\citep{vanderburg:2015,crossfield:2015a,montet:2015}.

There is an outside chance that transmission spectroscopy with JWST
could reveal the features of truly Earthlike atmospheres orbiting
late-type M dwarfs \citep{kaltenegger:2009,deming:2009}.  Such
measurements will be extremely challenging for even relatively
favorable planetary systems; whether these efforts will succeed depends on
JWST's on-orbit systematic noise floor, which is as yet unknown.
Depending on the success of JWST's transit spectroscopy, TESS and K2's
discoveries may justify development of dedicated instruments or
missions for dedicated spectroscopic surveys of large numbers of these
new systems
\citep{swain:2010thesis,swain:2012,glauser:2013,tinetti:2015,cowan:2015}.

The next generation of high-resolution, near-infrared spectrographs
\citep[see Table 1 in][]{crossfield:2014b} will measure abundances and
planetary fluxes at multiple longitudes for short-period planets
\citep[][]{dekok:2014}.  Together with low-resolution spectroscopic
phase curves from HST and JWST \citep{stevenson:2014c}, this means we
are entering the era of precision exocartography: such observations
will constrain temperature/pressure profiles and chemistry as a
function of planetary longitude.  From these observations, we can soon
expect {global temperature and chemical abundance maps for many
  exoplanets' atmospheres,} which will allow comparison to models of
atmospheric circulation and chemistry at an unprecedented level of
detail.

\subsection{The Longer Term: Beyond 2025}
A decade from now, JWST's extended mission will be near its end;
Spitzer, HST, Kepler, and TESS will be long defunct; today's
next-generation high-contrast imagers will have surveyed most of their
accessible systems. At this point the two most-anticipated
developments will be (1) the completion of one or more ground based,
giant segmented-mirror telescopes (GSMTs) operating in the optical and
infrared, and (2) NASA's planned WFIRST-AFTA mission, whose current
design includes a high-contrast imaging system operating at optical
wavelengths \citep{spergel:2015}.

The GSMTs will allow for instruments with significantly greater
spectral resolution and angular resolution.  With high-dispersion
spectrographs, GSMTs will measure atmospheric abundances and structure
and determine the signatures of atmospheric dynamics for many more
planets. Such instruments will also be capable of creating global
weather maps and movies of many brown dwarfs and some giant, young
planets on wide orbits via Doppler Imaging
\citep{crossfield:2014b,snellen:2014}.  These large telescopes might
even successfully detect O$_2$ or other biosignature gases using
high-dispersion spectroscopy on extrasolar planets orbiting small M
dwarfs \citep{snellen:2013a,rodler:2014}. Any detection would require
a dedicated campaign lasting many years \citep{rodler:2014}, but such
a result would capture the world's attention like never before.

The angular resolution provided by a GSMT's large aperture greatly
expands  the feasible direct-imaging science cases. That these will
include young, self-luminous planets has been long appreciated
\citep[e.g.,][]{macintosh:2006pfi}. Properly designed, these
ground-based imagers could access a small number of temperate
super-Earths, along with substantial numbers of larger planets already known
from radial velocity surveys \citep[][]{crossfield:2013a,quanz:2015}.
If high-dispersion spectroscopy is coupled to high-contrast
instruments, then the dual advantages of both high-contrast and high
spectral resolution would allow ELTs to probe yet fainter objects --
perhaps even habitable, Earth-sized planets orbiting nearby, bright M
dwarfs \citep{kawahara:2012,snellen:2015}.

A comparable but more sensitive instrument is the 2.4\,m-diameter
WFIRST-AFTA \citep{spergel:2015}.  Though still in its early stages,
current plans call for a visible-wavelength, low-resolution integral
field spectrograph capable of achieving planet/star contrast levels as
low as $10^{-10}$ (see Fig.~\ref{fig:contrast}). This facility would
enable a wide range of atmospheric science including reflected-light
spectroscopy of many known planets (down to smaller sizes than could
be accessed by a GSMT), and potentially even the discovery or
characterization of terrestrial planets in or near the habitable zone
around the nearest, most favorable FGK stars
\citep{marley:2014,burrows:2014b,traub:2014,greco:2015} --- complementing the
GSMT's high-contrast imaging around M dwarfs.

It is hoped that these ground- and space-based efforts will in turn
set the stage for a large-aperture, high-contrast, optical/infrared
mission in the coming decades \citep{kouveliotou:2014,tumlinson:2015}.
Aside from providing the best opportunity yet to survey worlds like
our own, such a program would facilitate an incredible diversity of
exoplanetary science and so continue to fuel the exoplanet revolution
for many years to come.



\acknowledgements{Acknowledgements}: 

I am grateful to J.~Lothringer, Prof.\ T.\ Barman, and Dr.\ E.\ Mills
for discussions and comments that substantially improved the quality of this
manuscript, and to Prof.\ L.\ Close and Dr.\ J.\ Males for useful
information about modern AO system performance. I also thank my
referee, Prof.\ J.\ Fortney, for hist extremely useful comments and
suggestions.  This work was performed under contract with the
California Institute of Technology/Jet Propulsion Lab funded by NASA
through the Sagan Fellowship Program executed by the NASA Exoplanet
Science Institute.

\bibliographystyle{apj}

\end{document}